\documentclass[twocolumn, times, twocolSection]{aastex631}
\usepackage{graphicx}
\usepackage{amsmath}
\usepackage{placeins}
\usepackage{multirow}
\usepackage{tablefootnote}
\usepackage{longtable}

\shorttitle{Simulating Earth Analogs for HWO and LIFE}
\shortauthors{Pradhan et al.}

\graphicspath{{./}{figures/}}

\begin{document}
	\title{Detectability of Atmospheric Biosignatures in Earth Analogs with Varying Surface Boundary Conditions: Prospects for Characterization in the UV, Visible, Near-Infrared, and Mid-Infrared Regions}

	\correspondingauthor{Liton Majumdar}
	\email{liton@niser.ac.in; dr.liton.majumdar@gmail.com}

	\author[0000-0003-4402-6475]{Dibya Bharati Pradhan}
	
	\affiliation{Exoplanets and Planetary Formation Group, School of Earth and Planetary Sciences, National Institute of Science Education and Research, Jatni 752050, Odisha, India}
	\affiliation{Homi Bhabha National Institute, Training School Complex, Anushaktinagar, Mumbai 400094, India}
	
	\author[ 0009-0002-4995-9346]{Priyankush Ghosh}
	\affiliation{Exoplanets and Planetary Formation Group, School of Earth and Planetary Sciences, National Institute of Science Education and Research, Jatni 752050, Odisha, India}
	\affiliation{Homi Bhabha National Institute, Training School Complex, Anushaktinagar, Mumbai 400094, India}

	\author[0009-0001-6483-7366]{Oommen P. Jose}
	\affiliation{Exoplanets and Planetary Formation Group, School of Earth and Planetary Sciences, National Institute of Science Education and Research, Jatni 752050, Odisha, India}
	\affiliation{Homi Bhabha National Institute, Training School Complex, Anushaktinagar, Mumbai 400094, India}
	
	\author[0000-0001-7031-8039]{Liton Majumdar}
	\affiliation{Exoplanets and Planetary Formation Group, School of Earth and Planetary Sciences, National Institute of Science Education and Research, Jatni 752050, Odisha, India}
	\affiliation{Homi Bhabha National Institute, Training School Complex, Anushaktinagar, Mumbai 400094, India}

	\begin{abstract}
		
		The search for potentially habitable exoplanets centers on detecting biosignature molecules in Earth-like atmospheres, which makes it essential to understand their detectability under biologically and geologically influenced conditions. In this study, we model the reflection and thermal emission spectra of such atmospheres across the UV/VIS/NIR and mid-IR regions and simulate their detectability with future mission concepts such as the Habitable Worlds Observatory (HWO) and the Large Interferometer for Exoplanets (LIFE). We employ Numerical Weather Prediction (NWP) model data, based on Earth's atmosphere, to derive temperature pressure profiles and couple them with a 1D photochemical model to assess the detectability of these molecules in Earth analogs located 10 parsecs away. We investigate the dominant reaction pathways and their contributions to the atmospheric composition of an Earth analog, with a focus on how they shape the resulting molecular signatures. We also examine the role of surface boundary conditions, which indirectly trace the effects of biological and geological processes, on the detectability of these molecules using HWO- and LIFE-type mission concepts. Our findings indicate that \(\textnormal{O}_3\) is detectable with both mission concepts, while \(\textnormal{H}_2\textnormal{O}\) requires specific surface humidity levels for detection with LIFE and shows only potential detectability with HWO. \(\textnormal{CO}_2\) is detectable with LIFE. Both \(\textnormal{N}_2\textnormal{O}\) and \(\textnormal{CH}_4\) require continuous surface outgassing for potential detection with LIFE, and \(\textnormal{CH}_4\) further requires low surface humidity to prevent masking by water features. Our work highlights the feasibility of characterizing the atmospheres of Earth analogs in the UV/VIS/NIR and mid-IR domains using HWO- and LIFE-type mission concepts and offers guidance for the development of future missions operating in these spectral regions.
		
	\end{abstract}
	

	\section{Introduction} \label{sec:intro}
	
	The discovery of over 6,000 exoplanets to date (NASA Exoplanet Archive\footnote{\url{https://exoplanetarchive.ipac.caltech.edu/}}) has significantly advanced the field of exoplanetary science, transforming our understanding of planetary systems beyond our own. This growing catalog of distant worlds underscores the importance of analyzing their atmospheres to gain insight into their underlying physical processes and chemical composition \citep{madhu19}. Observatories such as the Kepler Space Telescope, the Hubble Space Telescope (HST), and the James Webb Space Telescope (JWST) have played a pivotal role in revealing the extraordinary diversity of exoplanets \citep{seager2010exoplanet, sing2011hubble,fortney2024characterizing}. The Kepler mission, in particular, revolutionized our knowledge by identifying thousands of exoplanets and showcasing a wide variety of planetary types, from Earth-sized rocky worlds to gas giants, and exotic ``super-Earths" and ``mini-Neptunes" \citep{Kepler}. Hubble and Spitzer, with their space-based infrared photometry and spectrophotometry, provided the first glimpses into the compositions of exoplanets, detecting water vapor, clouds, and other key atmospheric components \citep{Charbonneau_2002, Charbonneau_2008}. More recently, JWST has furthered this exploration, offering unprecedented infrared sensitivity that enables scientists to study the atmospheres of smaller, cooler exoplanets in greater detail \citep{JWST1, JWST2}.
	
	As the field advances, the focus is gradually shifting towards discovering Earth-sized planets and characterizing temperate terrestrial worlds, with a major objective being the search for potential biological signatures, commonly referred to as ``biosignatures" in the astronomy and planetary science communities \citep{Robinson_2011_Epoxi, alei2024}. In the coming decades, efforts will be dedicated to refining detection methods to identify smaller, rocky planets within the habitable zones of their stars, where liquid water could exist. Research will also focus on studying the atmospheric properties of a large, statistically significant, and consistent observational sample of rocky exoplanets \citep{Fujii2018}. Direct imaging and spectroscopy are expected to be the primary methods to achieve this objective \citep{thayne_currie_2023}. Of all exoplanets discovered so far, only a handful have been successfully observed through direct imaging techniques \citep{zurlo_2024}. This stark contrast highlights the challenges of directly imaging exoplanets, which often require advanced instrumentation and favorable conditions. However, direct imaging is the only technique that allows for direct capture of photons emitted by exoplanets, setting it apart from other detection methods \citep{marois_2008b, Lagrange_2010, zurlo_2024}. Another way to directly image exoplanets, apart from using a coronagraph, is with a nulling interferometer \citep{Bracewell_1978, Angel_1997}. This technique cancels out the star's light and makes it easier to isolate the planet's thermal emission \citep{Quanz_2022a, quanz2022large}. A nulling interferometer helps achieve better spatial resolution. This allows for a detailed study of planetary systems that would be difficult to observe with other methods.
	
	Proposed future space-based mission concepts, such as the Large Interferometer for Exoplanets (LIFE)\footnote{\url{https://life-space-mission.com/}} and the Habitable Worlds Observatory (HWO)\footnote{\url{https://habitableworldsobservatory.org/home}}, offer the potential to target planetary thermal emission spectra \citep{quanz2018exoplanet,defrere2018,Fujii2018,lopezmorales2019,quanz2021atmospheric,quanz2022large,Alei_2022} and reflection spectra \citep{decadal}. The LIFE space observatory concept is primarily a nulling interferometer operating in the mid-infrared \citep{LIFE.I, quanz2018exoplanet, quanz2021atmospheric, quanz2022large}, designed to explore the atmospheric properties of a significant number of terrestrial exoplanets. LIFE's scientific focus aligns with the detection and characterization of temperate exoplanets in the mid-infrared, a potential topic for a future L-class mission in the ESA Science Program, as highlighted by ESA's Voyage 2050 Senior Committee report \footnote{\url{https://www.cosmos.esa.int/web/voyage-2050}}. On the other hand, the Astro2020 Decadal Survey has recommended that NASA develop a large (\(\sim6\, \text{m}\) inscribed diameter), stable, space-based infrared/optical/ultraviolet telescope capable of high-contrast imaging and spectroscopy as the next flagship observatory to search for biosignatures in habitable zone exoplanets, while simultaneously enabling transformative general astrophysics \citep{decadal}.
	
	\begin{figure*}[!ht]
		\centering
		\includegraphics[width=1\linewidth]{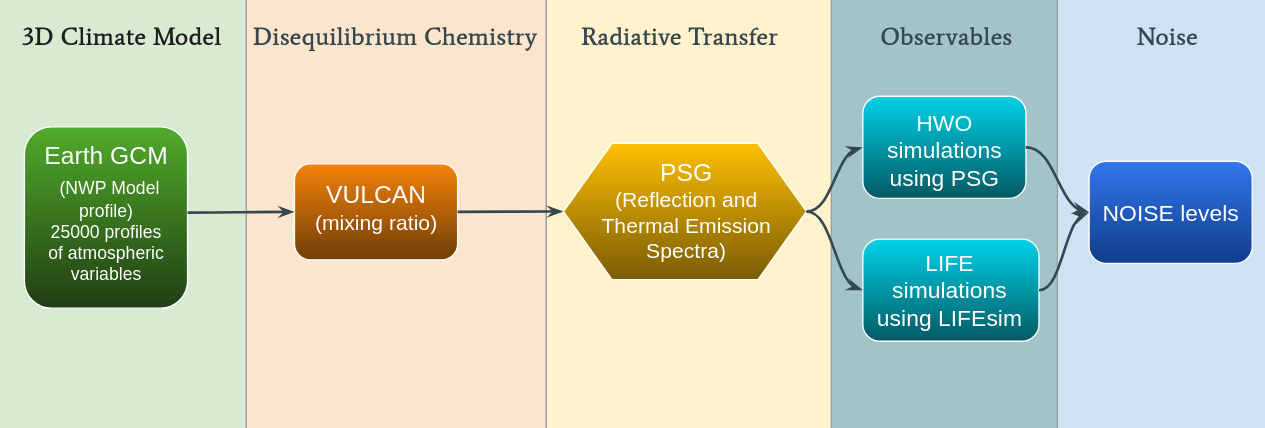}
		\caption{Flowchart representing our methodical pipeline. The model types are indicated in the large boxes: green for climate model, peach for chemistry model, yellow for radiative transfer model, cyan for observables simulator, and blue for noise output. These models are interconnected. The reflection spectra and thermal emission spectra represent the overall outputs of the pipeline, leading to the spectra that will be observed by future HWO and LIFE missions.}
		\label{fig:flowchart}
	\end{figure*}
	

The HWO and LIFE mission concepts focus on different wavelength regions, with HWO targeting the ultraviolet, visible, and near-infrared (UV/VIS/NIR) range of approximately 0.15 to 2 microns, while LIFE targets the mid-infrared (mid-IR) range of approximately 4 to 18.5 microns, primarily to study the atmospheres of temperate terrestrial exoplanets and provide complementary insights into planetary physics and chemistry. These missions collectively improve our ability to detect key biosignatures, refine planet characterization techniques, and produce complementary scientific results \citep{alei2024}. For example, reflected light observations provide information on a planet's dynamics, cloud composition, albedo, surface biosignatures, and ocean glint \citep{sagan_1993, Williams_2008, alei2024}. However, this method faces challenges, such as the strong correlation between radius and albedo in the presence of clouds, which complicates data interpretation. In contrast, thermal emission observations directly measure planetary radiation, offering insights into thermal structure and atmospheric composition, while being less affected by patchy clouds and providing more accurate radius measurements \citep{Alei_2022, alei2024}.

Several efforts have been made to study the reflection and thermal emission spectra of terrestrial exoplanets. Notably, the work of \citet{alei2024} stands out as the most recent study focusing on single-instrument and joint retrievals, highlighting both the HWO and LIFE concepts. The potential detection capabilities of a mission like LIFE have been well explored in studies by \citet{quanz2022large,Dannert2022,Kammerer2022}. The LIFE paper series has also extensively addressed the forward modeling \citep{Schwieterman_2022,schwieterman_Leung_2024,Angerhausen_2024} and retrieval of thermal emission spectra \citep{Alei_2022,konrad_2022,LIFE.IX}. However, there have not been many direct studies of HWO simulations considering a proper instrumental setup, except for a few \citet{alei2024,Young_2024}. The focus of \citet{Young_2024} is on the use of atmospheric retrieval models within a decision tree framework to analyze Earth-like exoplanets. The framework in the paper is designed to help identify biosignatures and planetary characteristics, specifically targeting modern Earth-like and Archean Earth-like analogs. \citet{Feng_2018} presents the observed planet-star flux ratio, considering an albedo model and integrating it with a direct imaging noise simulator while also conducting a retrieval study. Other notable retrieval studies include \citet{Damiano_2022,Robinson_Salvador_2023,Latouf_2023,Latouf_2024}, primarily focusing on Earth twin studies for the Habitable Exoplanet Observatory (HabEx) and the Large Ultraviolet Optical Infrared Surveyor (LUVOIR) concepts.

Earth is the best example of a habitable and inhabited world \citep{robinson2019earthexoplanet}. To identify a distant habitable world, it is important to model a planet as an Earth analog with varying surface boundary conditions to indirectly reflect active biological and geological processes, which play a fundamental role in regulating atmospheric composition. To understand the role of these mechanisms, we use data from the ECMWF-IFS model, a high-resolution Earth General Circulation Model (GCM) employed for numerical weather prediction (NWP), to obtain temperature-pressure (T-P) profiles. These profiles are then coupled with the 1D chemical kinetics model \texttt{VULCAN} to investigate the detectability of molecules influenced by surface boundary conditions. Our analysis focuses on their potential observability in the UV/VIS/NIR and mid-IR regions. We are using the scientific concepts laid out by the HWO and LIFE mission concepts as a foundation for our study of an Earth-analog located at a distance of 10 parsecs. This work would also help to consider the broader applicability to other future missions that aim to characterize terrestrial exoplanets in the UV, VIS, NIR, and mid-IR regions.

\begin{figure*}[!ht]
	\centering
	\includegraphics[width=\linewidth]{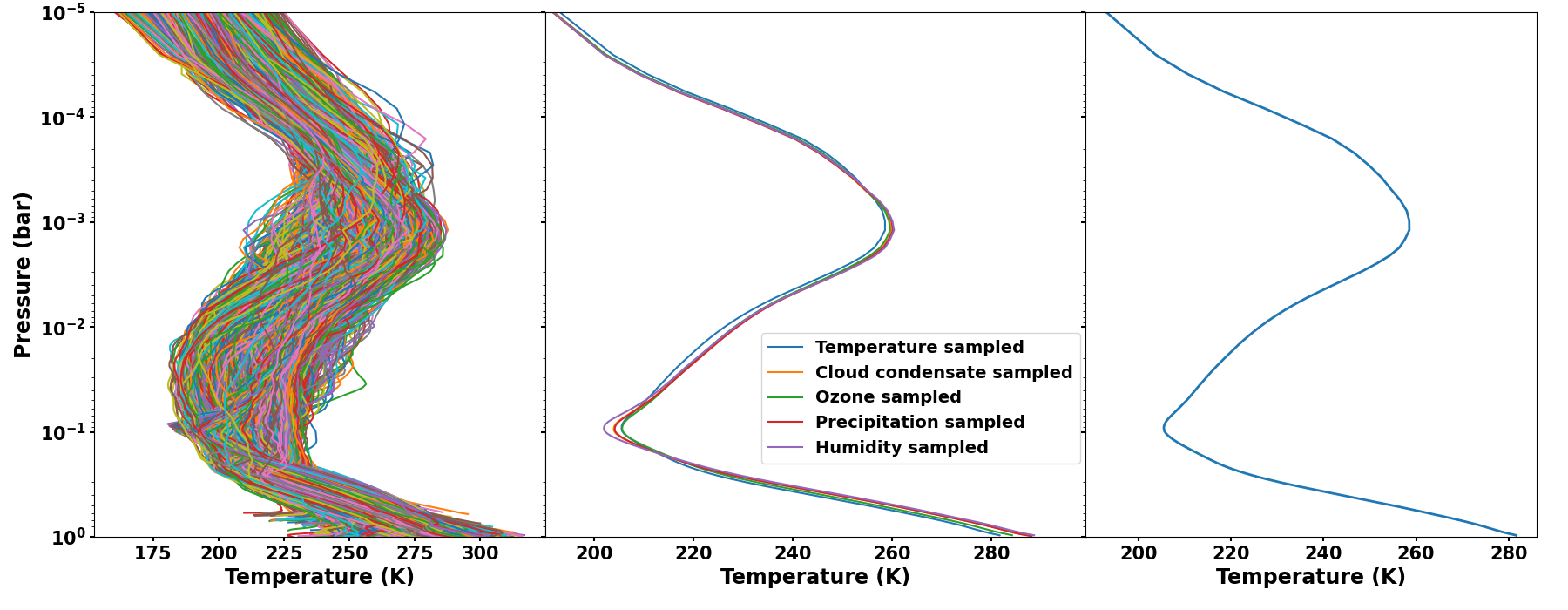}
	\caption{5000 temperature-pressure (T-P) profiles derived from a numerical weather prediction (\texttt{NWP}) model (left), an average T-P profile considering all five types of sampling techniques (middle), and the mean temperature-pressure profile representing the global average conditions of modern Earth (right). The leftmost panel illustrates the variability in T-P profiles due to different atmospheric conditions captured by the \texttt{NWP} model, while the far right panel provides a single, averaged T-P profile that can serve as a reference for global atmospheric studies of an Earth analog. The global average T-P profile remains the same across all our models: M1, M2, M3, M4, and M5.}
	\label{fig:TPplots}
\end{figure*}


The structure of the paper is as follows: In Section \ref{sec:meth}, we outline the workflow of the study, detailing the models used and the parameters adopted. The NWP model data, the chemical kinetics model with network updates, the approach taken for our simulations with different boundary conditions, and the models used for observation simulations, including their noise parameters, are all described here. The kinetics model output is presented in Section \ref{sec:res}, where we elaborate on the major chemical processes in the different layers of the atmosphere and present synthetic observations for all the simulations whose methodologies were covered in the previous section. The results are twofold: the reflection spectra to be obtained by HWO and the thermal emission spectra to be acquired by LIFE. We address the outputs of the simulations with various boundary conditions. We elaborate further on this topic in the section that follows, i.e., Section \ref{sec:dis}. We compare and explain the visible features and complementary nature of HWO and LIFE, along with the limitations of this study. Finally, our findings are drawn in Section \ref{sec:con}.

\section{Methods} \label{sec:meth}

To simulate Earth-like atmospheric conditions under disequilibrium chemistry, obtain both reflection and thermal emission spectra, and analyze the detectability of molecular features using the upcoming Habitable Worlds Observatory (HWO) and Large Interferometer for Exoplanets (LIFE) mission concepts, we have followed the methodology described in the flow chart illustrated in Figure \ref{fig:flowchart}. In this section, we provide a detailed description of the models used and the procedures followed, starting with the data from NWP (Section \ref{subsec: nwp}), followed by the chemical kinetic model (Section \ref{subsec: VULCAN}), and the reflection and thermal emission spectrum simulator (Section \ref{subsec: PSG}), along with the noise parameters (Section \ref{subsec: Noise}).

\subsection{Numerical Weather Prediction Model (NWP): Understanding the Thermal Structure} \label{subsec: nwp}

We obtain data for the temperature profiles, as shown in Figure \ref{fig:TPplots}, from the latest version of the atmospheric profile database of the forecasting system of the European Centre for Medium-range Weather Forecasts (ECMWF)\footnote{\url{https://nwp-saf.eumetsat.int/site/software/atmospheric-profile-data/}}. This data is generated from the Numerical Weather Prediction (\texttt{NWP}) model. Its Satellite Application Facility (\texttt{NWP SAF}) periodically produces datasets that contain sample profiles of atmospheric variables extracted from forcasting models. 

The latest version of the \texttt{NWP} model data comprises 25,000 atmospheric profiles sourced from globally operational short-range forecasts, typically covering time frames from a few hours to 3 days. This dataset captures normal conditions, typical variability, and extreme scenarios of atmospheric behavior on Earth. The short-range forecasts span from September 1, 2013, to August 31, 2014, and are generated four times daily at 00:00, 06:00, 12:00, and 18:00 UTC. The data is structured across 137 vertical pressure levels, ranging from the surface up to 0.01 hPa, with grid points spaced 16 km apart. These forecasts are produced by version Cy40r1 of the Integrated Forecasting System (IFS), which became operational at ECMWF on 19th November 2013. The IFS is a data assimilation system and a global numerical model for the Earth system. 

The IFS-137 database employs a sampling method very similar to the one described by \citet{chevallier_2006}. The sampling technique is based on maximizing the squared inter-profile departure, D, given by the equation:

\begin{equation}
	\mathrm{D} = \sum_{\mathrm{k}=1}^{\mathrm{K}} \sum_{\mathrm{m}=1}^{\mathrm{M}} \left(\frac{\mathrm{\theta}_{\mathrm{ik}}(\mathrm{m}) - \mathrm{\theta}_{\mathrm{jk}}(\mathrm{m})}{\mathrm{\sigma}_{\mathrm{k}}(\mathrm{m})}\right)^2
	\label{eq:departure}
\end{equation}

Here, $ \mathrm{k} $ and $ \mathrm{m} $ denote the indices for the variable and level, respectively. The values of variable $ \mathrm{k} $ at level $ \mathrm{m} $ in two different profiles are represented by $ \mathrm{\theta}_{\mathrm{ik}}(\mathrm{m}) $ and $ \mathrm{\theta}_{\mathrm{jk}}(\mathrm{m}) $. The variables and levels under consideration are denoted by $ \mathrm{K} $ and $ \mathrm{M} $, and $ \mathrm{\sigma}_{\mathrm{k}}(\mathrm{m}) $ represents the standard deviation of variable $ \mathrm{k} $ at level $ \mathrm{m} $. A profile is randomly selected from the available forecasts and stored in a separate list. The subsequent profile is then chosen from the forecast set and compared to the previously saved profile. If the deviation of the new profile exceeds that of the saved one, it replaces the previous profile in the list. To ensure a comprehensive representation of the input data without biasing towards the extremes, 90\% of the profiles in the dataset are selected randomly, as suggested by \citet{eresmaa_2014_nwp}.


This 25,000-profile database is divided into five subsets, with each subset containing 5000 profiles.
These five subsets are produced by sampling the corresponding five different parameters: temperature, specific humidity, ozone mixing ratio (sampled using univariate sampling by using $ \mathrm{K}=1 $ in Eq. (\ref{eq:departure})), cloud condensate, and precipitation (sampled using bivariate sampling by using $ \mathrm{K}=2 $ in Eq. (\ref{eq:departure})).

Out of the five sampling techniques used in the dataset, we are only considering the 5000 profiles that are temperature-sampled (shown in Figure \ref{fig:TPplots}) since we are only dealing with T-P profiles for our work. We also checked with all the other sampling techniques by taking an average T-P profile. However, we did not find any difference in the mean T-P profile, as represented in Figure \ref{fig:TPplots}. So, we have used the temperature-sampled-average T-P profile for our work out of all the other sampling techniques. The averaging is done by grouping the temperature data for each pressure level. The averaging is done to get a global average TP profile representative of Earth condition. Since the corresponding pressure values for each dataset are extremely close, the average of their corresponding temperature values remains approximately the same. This approximation is justified as each averaged (T, P) point is reasonably close to the subsequent one. The average T-P profile is shown in Figure \ref{fig:TPplots}. For the figures, the top of the atmosphere is taken at $10^{-5}$ bar and the bottom of the atmosphere at 1 bar. 

The global average T-P profile remains the same across all our simulations. Varying surface conditions are incorporated by modifying the lower boundary conditions in the 1D chemical-kinetics model \texttt{VULCAN}, which is discussed in Section \ref{subsec: VULCAN}, using this same averaged T-P profile.

\subsection{Chemical Kinetics Model VULCAN: Modelling the Evolution of Atmospheric Chemistry for Earth-like atmospheres under varying boundary conditions} \label{subsec: VULCAN}

\subsubsection{Model Configuration} \label{subsec:main_model}

To determine the chemical abundances corresponding to the mean temperature-pressure (T-P) profile derived from the \texttt{NWP} model (shown in Figure \ref{fig:TPplots}), we coupled the mean T-P profile with \texttt{VULCAN} 2.0\footnote{\url{https://github.com/exoclime/VULCAN}}, an open-source 1-D chemical kinetics code as described by \citet{Tsai_2021}. VULCAN solves the Eulerian continuity equation while incorporating the effects of condensation, advective transport, and chemical sources and sinks.

For our simulations, we considered 120 vertical layers, setting the pressure at the bottom to 1 bar and at the top to $5 \times 10^{-8}$ bar. This configuration provides sufficient vertical resolution to accurately capture atmospheric processes across a broad range of pressures. For the initial state of the atmosphere, we used constant mixing ratio values reflective of modern Earth's composition (listed in Table \ref{tab:abun}) which \texttt{VULCAN} utilizes to initiate chemical kinetics for modern Earth. The default solar flux for Earth, adopted in \texttt{VULCAN}, is based on a high-resolution spectrum from \citet{Gueymard_2018}.  We also activated condensation and settling processes to capture the tropospheric cold trap. Above this region, the abundance of water vapor is influenced by diffusion from the lower troposphere and methane oxidation, providing a more accurate representation of upper atmospheric dynamics \citep{Tsai_2021}. The other disequilibrium processes incorporated include eddy diffusion and molecular diffusion. The eddy diffusion is regulated by the eddy diffusion coefficient (K$_{zz}$) which is kept constant at $1\times 10^{5} \, \text{cm}^2 \text{s}^{-1}$ as it is the average of the K$_{zz}$ values considered for modern Earth in \citet{Tsai_2021}.

\begin{table}[!ht]
	\centering
	\caption{Initial atmospheric composition, detailing the constant mixing ratio values based on modern Earth composition used to initialize the \texttt{VULCAN} model.}
	\renewcommand{\arraystretch}{1.2} 
	\begin{tabular}{|c|c|}
		\hline \textbf{Species} & \textbf{Initial molecular mixing ratio}\\
		\hline
		$\mathrm{N}_2$ & $0.78$ \\
		$\mathrm{O}_2$ & $0.20$ \\
		$\mathrm{H}_2\mathrm{O}$ & $10^{-6}$ \\
		$\mathrm{CO}_2$ & $4\times 10^{-4}$ \\
		$\mathrm{Ar}$ & $9.34\times 10^{-3}$\\
		$\mathrm{SO}_2$ & $2\times 10^{-10}$\\
		\hline
	\end{tabular}
	\label{tab:abun}
\end{table}

\subsubsection{Adding reactions to the chemical network} \label{subsec: add_reactions}

For our analysis, we have used the \texttt{SNCHO Full Photo Network} which is the chemical network used in \citet{Tsai_2021} to benchmark modern Earth-like atmosphere. This network includes 96 species linked with 570 forward thermochemical reactions and 69 photodissociation branches. We extended this network by incorporating the reactions associated with $\mathrm{HO_2NO_2}$ listed in Table \ref{tab:reactions}. These reactions were taken from the chemical network used in \texttt{ATMOS}, a chemical kinetics code optimized for terrestrial exoplanets, as detailed by \citet{Arney_2014}.

\begin{table*}[!ht]
	\centering
	
	\caption{Reactions added to the SNCHO full photo chemical network for \texttt{VULCAN}.}
	\renewcommand{\arraystretch}{1.3} 
	\begin{tabular}{|lll|}
		\hline
		\textbf{Reaction} & \textbf{Rates} & \textbf{References}\\\hline
		$\mathrm{O + HO}_2\mathrm{NO}_2 \rightarrow \mathrm{OH} + \mathrm{NO}_2 + \mathrm{O}_2$    & $7.80\times 10^{-11} \text{exp}^{\frac{-3400}{T}}$                   &\citep{Chang_1981}\\
		$\mathrm{OH} + \mathrm{HO}_2\mathrm{NO}_2 \rightarrow \mathrm{H}_2\mathrm{O} + \mathrm{NO}_2 + \mathrm{O}_2$ & $1.3\times 10^{-12} \text{exp}^{\frac{380}{T}}$&\citep{Trevor_1988}, \citep{Smith_1984},   \\
		&&\citep{Barnes_1981}, \citep{Barnes_1982}\\
		$\mathrm{HO}_2 + \mathrm{NO}_2 + \mathrm{M} \rightarrow \mathrm{HO}_2\mathrm{NO}_2 + M$    & $1.9\times 10^{-31}\text{T}^{4\times 10^{-12}} \text{exp}^{\frac{-3.4}{T}}$ & IUPAC-04 \citep{IUPAC-04}\\
		$\mathrm{HO}_2\mathrm{NO}_2 + \text{h}\nu \rightarrow \mathrm{HO}_2 + \mathrm{NO}_2$& by photolysis cross section &SwRI\footnote{\url{https://phidrates.space.swri.edu/}\label{swri}} \citep{Huebner_Carpenter_1979}, \citep{Huebner_1992}\\
		$\mathrm{HO}_2\mathrm{NO}_2 + \text{h}\nu \rightarrow \mathrm{OH} + \mathrm{NO}_3$& by photolysis cross section &SwRI\textsuperscript{\ref{swri}} \citep{Huebner_Carpenter_1979}, \citep{Huebner_1992}\\
		\hline
		
	\end{tabular}
	
	\label{tab:reactions}
\end{table*}

These additional reactions primarily influence the photochemistry of the HO$_x$ and NO$_x$ families, particularly at higher altitudes or in regions characterized by lower temperatures. H$_2$O, NO$_2$, and O$_2$ are among the key molecules of focus in our study. Given that the added reactions involve the formation of these species, we anticipate that their chemistry may vary as a result of these reactions.


\subsubsection{The importance of surface boundary conditions and the initial composition}\label{subsec:boundary}

Boundary conditions are critical for understanding the planetary processes occurring near the surface and at the top of the atmosphere, as they play a significant role in shaping the atmospheric composition and dynamics. For the lower boundary conditions, sources and sinks are quantified by flux, mixing ratio, and (dry/wet) deposition velocities. Fluxes represent surface emission, while deposition rates dictate the removal of gases from the atmosphere at the lower boundary through processes such as gas absorption or uptake by the surface \citep{Hu_2012}. For rocky planets, constant fluxes are typically assumed. The dry deposition velocity \( \mathrm{v_d} \) of gases is calculated using the formula provided by \citet{wesley1989}:

\begin{equation}
	\mathrm{v_d} = \frac{-\mathrm{F}_c}{\mathrm{C}_z}
\end{equation}

Here, $\mathrm{F}_c$ is the flux density and $\mathrm{C}_z$ is the concentration at height $z$.

For our simulations, we consider surface emission and deposition rates for different species at the lower boundary conditions, utilizing the modern Earth benchmark from \citet{Tsai_2021}. This approach ensures that the model accurately reflects the influence of surface interactions on atmospheric chemical processes. In \texttt{VULCAN}, if the boundary condition for a particular molecule is not specified, a zero-flux condition is assumed \citep{Tsai_2021}.

\texttt{VULCAN}'s default surface boundary fluxes and deposition rates for modern Earth-like atmospheres include a variety of molecules. Among them, for \(\mathrm{N_2O}\) and \(\mathrm{CH_4}\), only outgassing flux is considered, with deposition set to zero. For \(\mathrm{H_2O}\) and \(\mathrm{CO_2}\), neither flux nor deposition is considered; instead, their mixing ratios are fixed at the surface. In addition to these default boundary conditions in \texttt{VULCAN}, we have introduced boundary values for \(\mathrm{CO_2}\) and \(\mathrm{O_2}\) in our simulations. The \(\mathrm{CO_2}\) flux and deposition rates are derived from \texttt{ATMOS} \citep{Arney_2014}. For \(\mathrm{O_2}\), we consider a flux value of \(7.1 \times 10^{9}\) molecules cm\(^{-2}\) s\(^{-1}\), which is derived from geochemical processes involving the subduction of \(\mathrm{Fe^{3+}}\) into the mantle over the past 4 billion years \citep{catling_2001, Kleinbohl_2018}. Since an established deposition rate for oxygen is unavailable in the literature, we assumed a value similar to that of \(\mathrm{CO_2}\), using \(1 \times 10^{-6}\) cm s\(^{-1}\).


\subsubsection{The Earth-like Models with varying boundary conditions}\label{subsec:test_models}

In our simulation, we investigated five distinct cases for Earth-like exoplanets, labeled as Models M1 through M5, as detailed in Table \ref{tab:atmospheric-models}. The primary model, described in Section \ref{subsec:main_model}, serves as the benchmark for present-day Earth, as outlined by \citet{Tsai_2021}. In this model, we consider the default boundary conditions with fixed surface mixing ratios for both $\mathrm{H_2O}$ and $\mathrm{CO_2}$.

The subsequent models, M1 through M5 vary in terms of their boundary conditions and whether additional surface fluxes and deposition velocities for $\mathrm{CO_2}$ and $\mathrm{O_2}$ are included. 
All these models have also been checked against 
the benchmarked modern Earth case in Section \ref{subsec:main_model}.

\begin{table*}[!ht]
	\centering
	\caption{Summary of Earth-like atmosphere models with constant initial atmosphere and varying boundary conditions.}
	\label{tab:atmospheric-models}
	\renewcommand{\arraystretch}{1.5} 
	\setlength{\tabcolsep}{10pt} 
	\begin{tabular}{|c|c|c|c|c|}
		\hline
		\textbf{Models} & \textbf{Composition of} & \multicolumn{2}{c|}{\textbf{Boundary Conditions}} & \textbf{Fixed species} \\ \cline{3-4}
		& \textbf{ atmosphere} & \textbf{Flux (cm\textsuperscript{-2} s\textsuperscript{-1})} & \textbf{v (dep)(cm s\textsuperscript{-1})} & \textbf{at surface} \\ \hline 
		M1 & \multirow{6}{*}{\parbox{3cm}{\textbf{Modern Earth-like:} \\ N\textsubscript{2}= 78\%, O\textsubscript{2}= 20\%, CO\textsubscript{2}= 0.04\%, H\textsubscript{2}O= $1 \times 10^{-4}$\%, Ar= 0.934\%, SO\textsubscript{2}= $2 \times 10^{-8}$\%}} & No Boundary & No Boundary & Nil \\ \cline{1-1} \cline{3-5}
		M2 & & No Boundary & No Boundary & H\textsubscript{2}O\footnote{H\textsubscript{2}O = 0.00894 \label{h20}}, CO\textsubscript{2}\footnote{CO\textsubscript{2} = $4 \times 10^{-4}$ \label{co2}} \\ \cline{1-1} \cline{3-5}
		M3 & & Default Boundary & Default Boundary & Nil \\ \cline{1-1} \cline{3-5}
		M4 & & O\textsubscript{2} = $7.1 \times 10^{9}$, CO\textsubscript{2} = $6.875 \times 10^{8}$ & O\textsubscript{2} = 0, CO\textsubscript{2} = $5 \times 10^{-5}$ & H\textsubscript{2}O\textsuperscript{\ref{h20}} \\ \cline{1-1} \cline{3-5}
		M5 & & O\textsubscript{2} = $7.1 \times 10^{9}$, CO\textsubscript{2} = $6.875 \times 10^{8}$ & O\textsubscript{2} = $1 \times 10^{-6}$, CO\textsubscript{2} = $5 \times 10^{-5}$ & H\textsubscript{2}O\textsuperscript{\ref{h20}} \\ \hline
	\end{tabular}
\end{table*}


For Models M1 and M2, we assessed the impact of maintaining fixed surface abundances for $\mathrm{CO_2}$ and $\mathrm{H_2O}$, excluding boundary values for other species. In \texttt{VULCAN}, the surface water vapor mixing ratio is set to 0.00894, which equates to 25\% relative humidity. Moreover, $\mathrm{CO_2}$ levels are fixed at 400 ppm since \texttt{VULCAN} does not factor in key processes that influence $\mathrm{CO_2}$ concentrations, such as biological activity, interactions with oceans, and geological processes \citep{Tsai_2021}. This led us to evaluate the sensitivity of these two species in Models M1 and M2. Next in model M3, we incorporated the default boundary values for other species, as presented in Table 2 of \citet{Tsai_2021}. However, unlike the main model, which fixes the mixing ratios for $\mathrm{H_2O}$ and $\mathrm{CO_2}$ at the surface, model M3 allows these mixing ratios to vary. Finally, for M4 and M5, we introduced $\mathrm{CO_2}$ boundary values from \texttt{ATMOS} \citep{Arney_2014} and flux for $\mathrm{O_2}$ from \citet{Kleinbohl_2018} as mentioned in Section \ref{subsec:boundary}. And, we then switched the sink (i.e., deposition velocity) on and off to assess its effect on the model outputs. A comprehensive analysis of the results from these models is provided in Section \ref{sec:apndx a} and \ref{sec:apndx c}.

\subsection{Planetary Spectrum Generator (PSG): Simulating Reflection and Thermal Emission Spectra} \label{subsec: PSG}

The radiance spectra for the mean temperature-pressure profile is generated by coupling the time-evolved chemical abundances from \texttt{VULCAN} to the radiative transfer suite Planetary Spectrum Generator (\texttt{PSG})\footnote{\url{https://psg.gsfc.nasa.gov/}} \citep{Villanueva2018psg}. 
It is a versatile and publicly accessible online radiative-transfer tool designed to simulate the remote spectral observables of planetary bodies across a wide range of viewing angles and distances \citep{Villanueva2018psg, Villanueva2022psg}. The PSG model uses the HITRAN 2020 database to obtain its input opacities \citep{Gordon2022}. In our study, \texttt{PSG} was employed to create reflection and thermal emission spectra based on our average temperature-pressure profile and corresponding chemical compositions outlined earlier. The parameters used to simulate reflection and thermal emission spectra for modern Earth-like exoplanets are provided in Table \ref{tab:psg_params}. The spectra were simulated for a wavelength range of 0.1 - 2 \textmu m for HWO simulations and 4 - 18.5 \textmu m for LIFE simulations. All simulations in this work involving PSG assume a phase angle of $77.8^\circ$, consistent with the approach used by \citet{alei2024}. Instead of modeling reflected light across the full planetary disk, we adopt a Lambertian surface and use a fixed average solar zenith angle of $77.756^\circ$ to approximate conditions near quadrature. This method provides realistic results with a single simulation, following the methodology described by \citet{alei2024}. All simulations in this study are performed under cloud-free conditions and assume a wavelength-independent surface albedo.

\subsection{Noise Calculations for HWO and LIFE} \label{subsec: Noise}

\subsubsection{\texttt{PSG}: Noise Calculations and Observation Simulation for HWO} \label{subsec: PSG noise SNR}

\texttt{PSG} includes a noise simulator capable of computing noise and sensitivity for a wide range of detectors and instrument configurations. In this work, we used \texttt{PSG} to simulate reflection spectra along with the expected noise levels detectable by HWO. The configuration parameters used to model noise for HWO are detailed in Table \ref{tab:psg_params2}. These instrument parameters are derived from the specifications outlined by \citet{decadal}. 

The considered wavelength range spans 0.15–2 \textmu m, with a telescope diameter of 6 meters. For noise calculations, we adopt a resolving power \( R \) of 7 between 0.15–0.51 \textmu m, \( R = 140 \) between 0.51–1 \textmu m, and \( R = 70 \) between 1–2.0 \textmu m for HWO, in accordance with values from the LUVOIR-B concept as discussed by \citet{alei2024}. We assume an observation time of 100 hours to simulate the reflection spectra for all our models. 

To benchmark our noise simulation for HWO against \citet{alei2024}, we scaled the noise and calibrated the reflected spectra to achieve a target signal-to-noise ratio (S/N) across three wavelength intervals: 0.15–0.51~\textmu m (UV), 0.51–1~\textmu m (VIS), and 1–2~\textmu m (NIR). Specifically, we ensured an S/N of 10 at three reference wavelengths, 0.35~\textmu m (UV), 0.55~\textmu m (VIS), and 1.2~\textmu m (NIR), and scaled the S/N across the remaining spectral range accordingly. Additionally, we generated 1$\sigma$ and 2$\sigma$ noise regions to represent the uncertainty across the spectrum. 

As shown in Figure \ref{fig:HWO_final_plot}, the magnitude of the flux and the corresponding noise closely match the results of \citet{alei2024}, demonstrating excellent agreement.

\begin{table}[!ht]
	\centering
	\caption{Parameters that were used in \texttt{PSG} for generating emission and reflection spectra. The stellar and planetary parameters were taken to represent a modern Earth analog around a sun-like star.}
	\label{tab:psg_params}
	\renewcommand{\arraystretch}{1.4} 
	\resizebox{\columnwidth}{!}{%
		\begin{tabular}{|lc|}
			\hline
			\multicolumn{1}{|c|}{\textbf{PSG parameters}} & \textbf{Emission \& Reflection spectra} \\ \hline
			\multicolumn{2}{|l|}{\textbf{Stellar parameters}}                                       \\ \hline
			\multicolumn{1}{|l|}{Star temperature}        & 5778 K                                  \\
			\multicolumn{1}{|l|}{Radius}                  & 1 $R_\sun$                                \\
			\multicolumn{1}{|l|}{Semi-major axis}         & 0.9837 AU                               \\
			\multicolumn{1}{|l|}{Star type}               & G-type                                  \\ \hline
			\multicolumn{2}{|l|}{\textbf{Planetary parameters}}                                     \\ \hline
			\multicolumn{1}{|l|}{Radius}                  & 1 $R_\earth$                              \\
			\multicolumn{1}{|l|}{Distance}                & 10 pc                                    \\
			\multicolumn{1}{|l|}{Phase}                   & $77.8^\circ$                                    \\ \hline
			\multicolumn{2}{|l|}{\textbf{Surface and Atmospheric parameters}}                       \\ \hline
			\multicolumn{1}{|l|}{Surface pressure}        & 1013 mbar                               \\
			\multicolumn{1}{|l|}{Surface temperature}     & 288 K                                   \\
			\multicolumn{1}{|l|}{Albedo}                  & 0.3                                     \\
			\multicolumn{1}{|l|}{Emissivity}              & 0.692                                   \\
			\multicolumn{1}{|l|}{Mean mol. weight}        & 28.97 g/mol                             \\ \hline
		\end{tabular}%
	}
\end{table}

\begin{table}[!ht]
	\centering
	\caption{Noise parameters considered for simulations for HWO using \texttt{PSG}.}
	\label{tab:psg_params2}
	\renewcommand{\arraystretch}{1.2} 
	\resizebox{\columnwidth}{!}{%
		\begin{tabular}{|l|c|}
			\hline
			\textbf{PSG parameters}   & \multicolumn{1}{l|}{\textbf{HWO simulations}} \\ \hline
			Aperture diameter         & 6 m                                           \\
			Spectral resolution       & 7, 140, 70                                    \\
			Wavelength range          & 0.15 \textmu m -- 2 \textmu m                             \\
			Beam (FWHM)               & 0.045 arcsec                                  \\
			Exozodi                   & 4.5 local zodi                                \\
			Contrast                  & $10^{-10}$                                       \\
			Exposure                  & 3600 secs                                     \\
			Number of exposures       & 100                                           \\
			Number of pixels          & 10                                            \\
			Temperature of the optics & 273 K                                         \\
			Emissivity of the optics  & 0.1                                           \\
			Inner working angle (IWA) & 3.73 $\lambda$/D                              \\ \hline
		\end{tabular}%
	}
\end{table}

\subsubsection{\texttt{LIFEsim}: Noise Calculations and Observation Simulation for LIFE} \label{subsec: LIFEsim noise SNR}

\texttt{LIFEsim} is a state-of-the-art simulator specifically designed for the LIFE mission concept, providing accurate and reliable simulations of the interferometric measurement processes involved in its operation. This tool enables us to assess the capabilities of LIFE and explore the parameter space for the detection and characterization of exoplanets \citep{Dannert2022}. In our study, we input the theoretical thermal emission spectrum obtained from \texttt{PSG} into \texttt{LIFEsim} to generate noise and simulate observations for LIFE.

Since our study focuses on temperate terrestrial ``Earth-like" planets, we assume the planet's radius, surface gravity, and angular separation to be the same as those of Earth in the Sun-Earth system (R = 6378.14 km, ang. sep. = 0.1 arcsec, and g = 9.8 m s$^{-2}$). The star-planet system is simulated at a distance of 10 parsecs. The simulation parameters used in \texttt{LIFEsim} are detailed in Table \ref{tab:noise_life}. 

We consider \texttt{LIFEsim} in its ``baseline" setup, which includes four collector apertures, each with a diameter of 2 m, a broadband wavelength range of 4 to 18.5 \textmu m, a throughput of 5\%, and a spectral resolution of R = 50 \citep{quanz2022large, Dannert2022}. For the noise parameters, we adopt the baseline resolution of the instrument.  All the configuration parameters for noise, except for the exozodiacal dust level and spectral resolution, align with previous studies on LIFE. We assume the exozodiacal dust level to be 4.5 times the local zodiacal dust level, consistent with the approach taken by \citet{alei2024}.

\citet{Angerhausen_2024} calculated how long certain molecular features (such as $\mathrm{N_2O}$ and $\mathrm{CH_3X}$) in the thermal emission spectra of an Earth-like exoplanet around a Sun-like star at a distance of 5 pc would take to become distinguishable by varying the integration periods. They concluded that a 10-day observation period is relatively short, making it suitable for characterizing a larger sample of planets. A 50-day observation allows for a more thorough analysis, while a 100-day observation is optimistic and primarily applicable to the most promising targets for LIFE. Thus, in our study, we modeled the detectability of different molecules by considering a fixed observation duration of 50 days—an intermediate case from \citet{Angerhausen_2024}—which also closely matches the required observation time for an Earth-twin at 10 pc for a nominal setup of R = 50, S/N = 10, and a 2m aperture diameter, as presented in Table 6 of \citet{konrad_2022}.

To benchmark our noise simulation for LIFE against \citet{alei2024}, we  consider a fixed S/N of 10 at 11.2 \textmu m, for the thermal emission spectra. As shown in Figure \ref{fig:LIFE_final_plot}, the magnitude of the flux and the corresponding noise closely match the results of \citet{alei2024}, demonstrating excellent agreement for LIFE noise calculations as well.

\subsubsection{Metric for Molecular Detectability in HWO and LIFE} \label{subsubsec:BandSNR}

For our analysis, we assess the detectability of spectral features using a key metric: the band-integrated signal-to-noise ratio (SNR), computed following the approach of \citet{Angerhausen_2024}.

	\begin{equation}
		\text{SNR} = \sqrt{\sum_{i=1}^{n} \left( \frac{\Delta y_i}{\sigma(y_i)} \right)^2 }
		\label{eq:snr}
	\end{equation}

Here, $\Delta y_i$ denotes the flux difference between two spectra: one including the target spectral feature and the other excluding it, for each of the $n$ spectral bins. The term $\sigma(y_i)$ represents the instrumental noise in the same bin, determined from the observational sensitivity. For this noise term, we use the actual noise output from PSG for HWO spectra without any rescaling, and the corresponding noise values provided by the \texttt{LIFEsim} simulator for LIFE spectra. The resulting band-integrated SNR values for all molecular features are listed in Table~\ref{tab:Band_SNR_HWO_LIFE}, and the related analysis is discussed in Section~\ref{sec:dis}.

As prescribed by \citet{Angerhausen_2024}, for N$_2$O and certain other molecules in LIFE spectra, considering an Earth-like planet 5 pc away orbiting a Sun-like star, a spectral feature is considered distinguishable if the band-integrated SNR exceeds a threshold of $5$–$10$. In this work, for a planet located 10 pc away around a Sun-like star, we tabulate the band-integrated SNR for different molecular features for HWO and LIFE. A molecule is considered detectable if the band-integrated SNR meets or exceeds the cited threshold range. Throughout this work, we adopt the following detectability convention for the band-integrated SNR (see equation~\ref{eq:snr}): Detectable if SNR $\geq 10$, Potentially or marginally detectable if $5 \leq \text{SNR} < 10$, and Not detectable if SNR $< 5$. All detection statements are conditional on the assumed instrument setups and integration times.

\begin{table}[!ht]
	\centering
	\caption{Parameters considered for simulations for LIFE using \texttt{LIFEsim}. 
		For detailed explanations of the parameters, refer to \citep{Dannert2022}.}
	\label{tab:noise_life}
	\renewcommand{\arraystretch}{1.2} 
	\resizebox{\columnwidth}{!}{%
		\begin{tabular}{|l|c|}
			\hline
			\textbf{LIFEsim parameters}   & \multicolumn{1}{l|}{\textbf{LIFE simulations}} \\ \hline
			Collector aperture diameter   & 2 m                                            \\
			Spectral resolution           & 50                                             \\
			Wavelength range              & 4 \textmu m -- 18.5 \textmu m                    \\
			Eclliptic latitude            & 0.78 rad                                       \\
			Exozodi                       & 4.5 local zodi                                 \\
			Planet radius                 & 1 $R_\earth$                                     \\
			Star radius                   & 1 $R_\sun$                                       \\
			Target distance               & 10 pc                                           \\
			Stallar effective temperature & 5778 K                                         \\
			Angular separation            & 0.1 arcsec                                     \\
			Array baseline                & optimized for planet                           \\
			Integration time              & 50 days                                        \\
			Quantum efficiency            & 70\%                                           \\
			Photon throughput             & 5\%                                            \\
			Nulling baseline length       & 10-100 m                                        \\
			Array baseline ratio          & 6:1                                            \\
			Slew rate                     & 10 hr                                            \\
			Time efficiency               & 80\%                                           \\ \hline
		\end{tabular}%
	}
\end{table}

\vspace{-1.0em}
\section{Results} \label{sec:res} 
This section presents the output from the chemical kinetics model discussed in Section \ref{subsec: VULCAN} and highlights the major chemical processes occurring in different atmospheric layers for a modern Earth-like benchmark. Following this, we discuss the simulations of the reflection spectra in the UV/VIS/NIR region for HWO in Section \ref{subsec:reflection}, as well as the thermal emission spectra in the mid-IR region for LIFE in Section \ref{subsec:thermal_emission}, which are benchmarked against \citet{alei2024}. Finally, we examine different Earth-like model simulations, introduced in Table \ref{tab:atmospheric-models}, and their corresponding chemistry and spectra in Sections \ref{sec:apndx a} and \ref{sec:apndx c}.

\subsection{Atmospheric Chemistry: Depicting Chemical Processes using VULCAN} \label{subsec: chemistry}

\begin{figure}[!ht]
	\centering
	\includegraphics[width=1\linewidth]{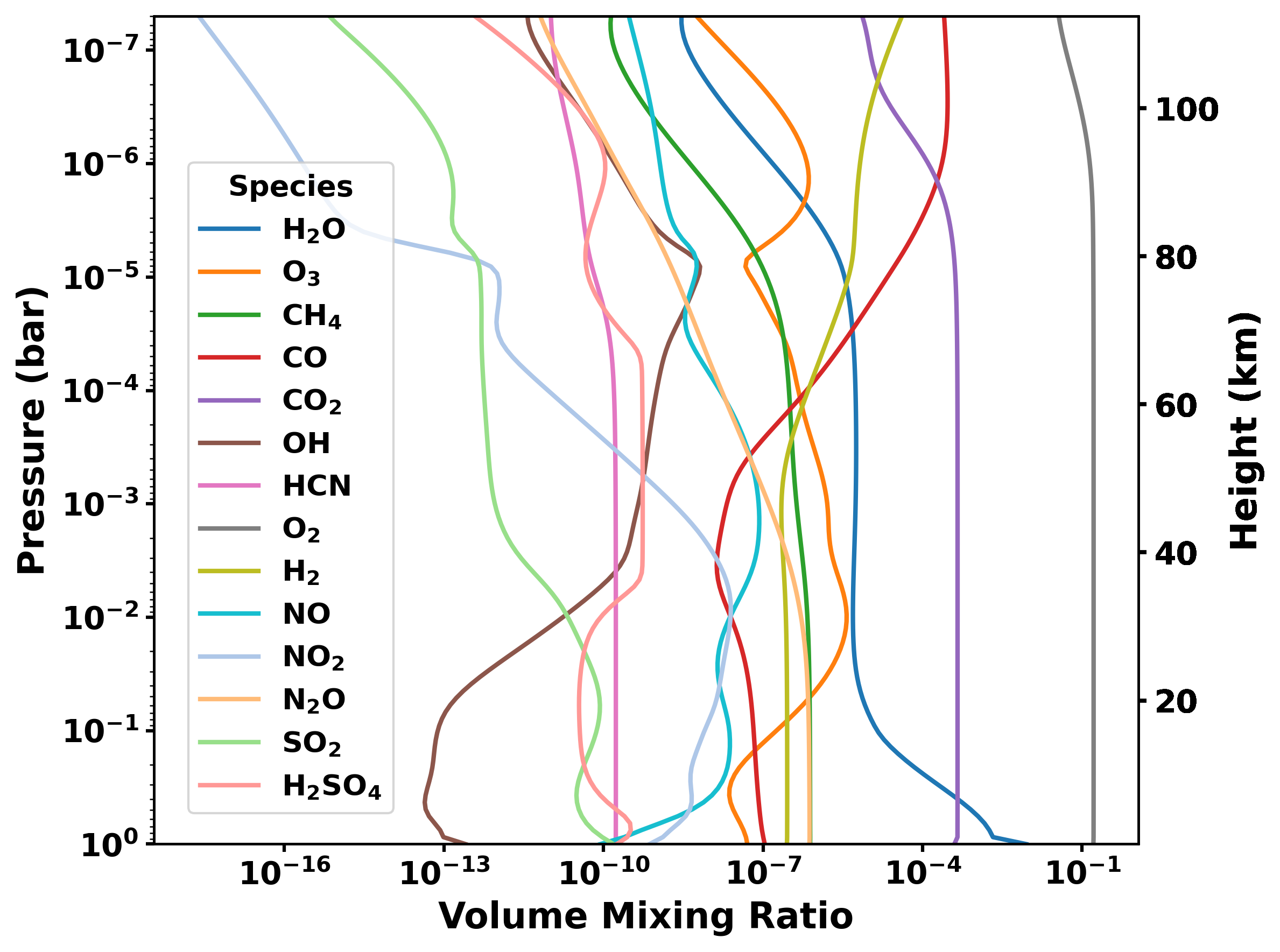}
	\caption{Mixing ratio profiles obtained from VULCAN for the temperature sampled average T-P profile. The molecules are $\mathrm{H_2O}$, $\mathrm{O_3}$, $\mathrm{CH_4}$, $\mathrm{CO}$, $\mathrm{CO_2}$, $\mathrm{OH}$, $\mathrm{HCN}$, $\mathrm{O_2}$, $\mathrm{H_2}$, $\mathrm{NO}$, $\mathrm{NO_2}$, $\mathrm{N_2O}$, $\mathrm{SO_2}$, $\mathrm{H_2SO_4}$. The molecules are selected considering their average mixing ratios $> 10^{-18}$  between 1 bar and 0.01 mbar.}
	\label{fig:mix_ratio_avg}
\end{figure}

Disequilibrium chemical processes play a major role in shaping the atmospheric composition of modern Earth. Our chemical kinetics model simulations capture the effects of these processes on the final evolved abundances.
\par
Figure \ref{fig:mix_ratio_avg} presents the mixing ratio profiles simulated by \texttt{VULCAN} for the average temperature structure of modern Earth's atmosphere, derived from \texttt{NWP} model data (see Section \ref{subsec: nwp}). The species that play a crucial role in atmospheric chemistry and constitute the bulk composition of the atmosphere have volume mixing ratios (VMR) $\gtrsim 10^{-8}$ and serve as primary observable targets. To determine which chemical reactions significantly contribute to observable features in emission and reflection spectra, we examine different atmospheric layers from the surface to the uppermost regions.
\par
We analyze all reactions across all atmospheric layers for molecules that exhibit observable features in the HWO and LIFE spectra.

To quantify the contribution of each major production and destruction reaction for every species, we calculate the percentage contribution of each reaction. The production or destruction percentage contribution of any reactant or product in a two-body reaction, such as  
	$\mathrm{A} + \mathrm{B} \rightarrow \mathrm{C} + \mathrm{D}$,  
	a three-body reaction,  
	$\mathrm{A} + \mathrm{B} + \mathrm{M} \rightarrow \mathrm{C} + \mathrm{M}$,  
	or a photochemical reaction,  
	$\mathrm{A} + \mathrm{h\nu} \rightarrow \mathrm{C} + \mathrm{D}$,  
	is given by:

	\begin{equation}
		r_{\mathrm{d,p}} = \frac{k_{\mathrm{2b,3b,ph}} \times \prod_{i} n_{\mathrm{i}} }{\sum_{d,p} k_{\mathrm{2b,3b,ph}} \times \prod_{i} n_{\mathrm{i}} } \times 100 \%
\end{equation}

Here, $k_{\mathrm{2b,3b,ph}}$ represents the rate coefficient of the two-body (2b), three-body (3b), or photochemical (ph) reaction under consideration, and $n_{i}$ denotes the number density of the $i^{\mathrm{th}}$ reactant in that reaction. The product is taken over all reactants, and the summation is performed over all reactions that destroy or produce the species of interest.  
	In terms of volume mixing ratios, the number density is expressed as $n_i = (\mathrm{VMR})_i \times n_{\mathrm{gas}}$. The total gas number density $n_{\mathrm{gas}}$ for a given atmospheric layer is determined from the ideal gas law:  
	$n_{\mathrm{gas}} = p / (k_B T)$,  
	where $p$ is the pressure, $k_B$ is the Boltzmann constant, and $T$ is the temperature of the layer.

To assess the dominant formation and destruction pathways of the major contributors to the observed reflected and emitted fluxes on modern Earth-like exoplanets, we identify reactions contributing more than 10\% to the total production or destruction of each species at any altitude. These key reactions and their relative contributions are summarized in Table~\ref{tab:chem_rxns} for both HWO and LIFE. For each molecule, the “\% contribution” column lists the local maxima and minima of each reaction’s fractional contribution to its total production or destruction rate, with the corresponding altitudes shown in the adjacent column. 

Note that the percentages do not sum to 100\%, since each value corresponds to a local maximum or minimum specific to an individual reaction and does not represent a complete partitioning of the destruction/production pathways at a given altitude. For example, the first line in the table represents a major destruction pathway of $\mathrm{O_3}$. As altitude increases from 0~km to 4.5~km, the percentage contribution of this reaction to the total $\mathrm{O_3}$ destruction decreases from 73\% to 40.15\%. Between 4.5~km and 31.32~km, the contribution increases again from 40.15\% to a maximum of 47.41\%, before declining to 10.36\%. 

In some cases, such as the production of $\mathrm{N_2O}$ via $\mathrm{NH_2}$ and $\mathrm{NO_2}$, a constant dominance of nearly 100\% persists from the surface up to 30~km, after which the contribution begins to decline. For all reactions, we report contributions down to $\sim$10\%. If the percentage rises above $\sim$10\% again at higher altitudes, those values are shown as well. 

At any given altitude, the total percentage contribution of all destruction as well as production reactions of any molecule (e.g., $\mathrm{O_3}$) will sum to 100\%. We have not aligned percentage values across the same altitudes for all molecules, since a given altitude may not correspond to a local maximum or minimum for every species.


\startlongtable
\begin{deluxetable*}{c c c c c}
	\tablecaption{Major chemical reaction pathways and their contributions to the atmosphere of modern Earth-like exoplanets.\label{tab:chem_rxns}}
	\tablehead{
		\hline
		\colhead{\textbf{Species}} & \colhead{\textbf{Reaction}} & \colhead{\textbf{Reaction Type}} & \colhead{\textbf{\% Contribution}} & \colhead{\textbf{Altitude (km)}} \\
		\hline
	}
	\startdata
	$\mathrm{O_3}$ & $\mathrm{O_3 \rightarrow O_2 + O}$\tablenotemark{a} & Photo-dissociation & 73 -- 40.15 -- 47.41 -- 10.36 & 0 -- 4.5 -- 31.32 -- 79.11 \\
	& $\mathrm{NO + O_3 \rightarrow NO_2 + O_2}$\tablenotemark{a} & Two-body & 14.68 -- 76 -- 5.26 & 0 -- 8.76 -- 35.19 \\
	& $\mathrm{O_3 \rightarrow O_2 + O(^1D)}$\tablenotemark{a} & Photo-dissociation & 11.68 -- 10.13 & 0 -- 3.47 \\
	& & & 11.11 -- 85.33 -- 25.81 -- 31.25 & 16.06 -- 58.61 -- 89.85 -- 112.9 \\
	& $\mathrm{O_3 + H \rightarrow OH + O_2}$\tablenotemark{a} & Two-body & 11.57 -- 69.58 -- 62.9 & 75.75 -- 89.85 -- 112.9 \\
	& $\mathrm{O + O_2 + M \rightarrow O_3 + M}$\tablenotemark{b} & Three-body & 100 & 0 -- 112.9 \\
	& & & & \\
	$\mathrm{O_2}$ & $\mathrm{O + O_2 + M \rightarrow O_3 + M}$\tablenotemark{a} & Three-body & 96.08 -- 98.8 -- 15.32 -- 63.03 -- 13 & 0 -- 8.76 -- 81.59 -- 90.68 -- 100 \\
	& $\mathrm{O(^1D) + O_2 \rightarrow O + O_2}$\tablenotemark{a} & Two-body & 10.25 -- 16.37 -- 10.06 & 32.28 -- 52.19 -- 74.03 \\
	& $\mathrm{O_2 \rightarrow O + O}$\tablenotemark{a} & Photo-dissociation & 10.1 -- 41.85 -- 24.94 -- 41.82 -- 24.7 & 70.47 -- 81.59 -- 89.02 -- 98.28 -- 112.9 \\
	& $\mathrm{H + O_2 + M \rightarrow HO_2 + M}$\tablenotemark{a} & Three-body & 10.63 -- 38.82 -- 13.34 & 72.27 -- 82.41 -- 86.54 \\
	& $\mathrm{O_2 \rightarrow O + O(^1D)}$\tablenotemark{a} & Photo-dissociation & 10.78 -- 70.06 & 94.87 -- 112.9 \\
	& $\mathrm{O_3 \rightarrow O_2 + O}$\tablenotemark{b} & Photo-dissociation & 70.95 -- 18.51 -- 58.01 -- 10.08 & 0 -- 8.76 -- 23.94 -- 67.68 \\
	& $\mathrm{NO + O_3 \rightarrow NO_2 + O_2}$\tablenotemark{b} & Two-body & 14.27 -- 75.72 -- 11.04 & 0 -- 8.76 -- 25.75 \\
	& $\mathrm{O_3 \rightarrow O_2 + O(^1D)}$\tablenotemark{b} & Photo-dissociation & 11.35 -- 10.72 & 0 -- 2.33 \\
	& & & 10.83 -- 69.75 -- 12.47 & 16.06 -- 52.19 -- 79.94 \\
	& & & 12.88 -- 10.1 & 90.68 -- 95.72 \\
	& $\mathrm{O(^1D) + O_2 \rightarrow O + O_2}$\tablenotemark{b} & Two-body & 10.27 -- 16.42 -- 10.07 & 32.28 -- 52.19 -- 74.03 \\
	& & & 10.05 -- 70.78 & 100.02 -- 112.9 \\
	& $\mathrm{O + OH \rightarrow O_2 + H}$\tablenotemark{b} & Two-body & 10.12 -- 46.12 -- 11.07 & 70.47 -- 82.41 -- 101.78 \\
	& $\mathrm{O + HO_2 \rightarrow OH + O_2}$\tablenotemark{b} & Two-body & 10.59 -- 37.14 -- 11.05 & 72.27 -- 81.59 -- 86.54 \\
	& $\mathrm{O_3 + H \rightarrow OH + O_2}$\tablenotemark{b} & Two-body & 11.29 -- 34.89 -- 10.08 & 83.23 -- 89.85 -- 101.78 \\
	& $\mathrm{O + O + M \rightarrow O_2 + M}$\tablenotemark{b} & Three-body & 10.13 -- 58.9 -- 26.14 & 91.51 -- 103.57 -- 112.9 \\
	& & & & \\
	$\mathrm{H_2O}$ & $\mathrm{O(^1D) + H_2O \rightarrow OH + OH}$\tablenotemark{a} & Two-body & 97.38 -- 82.88 -- 85.12 -- 12.88 & 0 -- 35.19 -- 47.83 -- 73.16 \\
	& $\mathrm{H_2O \rightarrow H + OH}$\tablenotemark{a} & Photo-dissociation & 10.11 -- 17.08 -- 14.84 -- 88.42 -- 85.53 & 23.94 -- 35.19 -- 47.83 -- 101.78 -- 112.9 \\
	& $\mathrm{HNO_3 + OH \rightarrow NO_3 + H_2O}$\tablenotemark{b} & Two-body & 46.4 -- 91.34 -- 73.52 -- 85.95 -- 11.27 & 0 -- 16.06 -- 23.04 -- 28.5 -- 40.28 \\
	& $\mathrm{OH + CH_4 \rightarrow H_2O + CH_3}$\tablenotemark{b} & Two-body & 17.97 -- 10.8 & 0 -- 7.76 \\
	& & & 10.3 -- 13.1 -- 10.21 & 35.19 -- 39.24 -- 43.48 \\
	& $\mathrm{NO + NH_2 \rightarrow N_2 + H_2O}$\tablenotemark{b} & Two-body & 11.03 -- 13 -- 10.45 & 20.39 -- 22.15 -- 23.94 \\
	& $\mathrm{OH + HO_2 \rightarrow H_2O + O_2}$\tablenotemark{b} & Two-body & 11.67 -- 96.42 -- 67.86 -- 13.94 & 33.24 -- 65.75 -- 80.76 -- 84.06 \\
	& $\mathrm{HO_2 + H \rightarrow O + H_2O}$\tablenotemark{b} & Two-body & 11.09 -- 30.13 -- 99.35 & 76.6 -- 80.76 -- 111.97 \\
	& & & & \\
	$\mathrm{CO_2}$ & $\mathrm{O(^1D) + CO_2 \rightarrow O + CO_2}$\tablenotemark{a} & Two-body & 100 -- 11.09 & 0 -- 91.51 \\
	& & & 10.57 -- 14.59 & 109.1 -- 112.9 \\
	& $\mathrm{CO_2 \rightarrow CO + O(^1D)}$\tablenotemark{a} & Photo-dissociation & 12.02 -- 49.5 -- 10.5 & 77.44 -- 82.41 -- 90.68 \\
	& $\mathrm{N + CO_2 \rightarrow CO + NO}$\tablenotemark{a} & Two-body & 16.97 -- 92.28 -- 65.49 & 86.54 -- 97.42 -- 112.9 \\
	& $\mathrm{OH + CO \rightarrow H + CO_2}$\tablenotemark{b} & Two-body & 95.48 -- 12.5 & 0 -- 15.19 \\
	& & & 10.05 -- 89.23 -- 10.55 & 70.47 -- 81.59 -- 88.19 \\
	& $\mathrm{O(^1D) + CO_2 \rightarrow O + CO_2}$\tablenotemark{b} & Two-body & 16.4 -- 99.83 -- 10.52 -- 79.8 -- 10.08 & 2.33 -- 51.1 -- 81.59 -- 89.02 -- 109.1 \\
	& $\mathrm{O(^1D) + CO \rightarrow CO_2}$\tablenotemark{b} & Two-body & 12.2 -- 91.66 & 89.02 -- 112.9 \\
	& & & & \\
	$\mathrm{CH_4}$ & $\mathrm{OH + CH_4 \rightarrow H_2O + CH_3}$\tablenotemark{a} & Two-body & 99.98 -- 43.97 -- 80.43 -- 12.25 & 0 -- 23.94 -- 40.28 -- 71.38 \\
	& $\mathrm{O(^1D) + CH_4 \rightarrow OH + CH_3}$\tablenotemark{a} & Two-body & 13.2 -- 52.51 -- 18.34 -- 54.26 -- 13.7 & 16.06 -- 23.94 -- 40.28 -- 67.68 -- 73.16 \\
	& $\mathrm{CH_4 \rightarrow ^1CH_2 + H_2}$\tablenotemark{a} & Photo-dissociation & 13.69 -- 47.44 -- 44.25 & 70.47 -- 82.41 -- 111.97 \\
	& $\mathrm{CH_4 \rightarrow CH_3 + H}$\tablenotemark{a} & Photo-dissociation & 12.02 -- 41.64 -- 36.39 & 70.47 -- 83.23 -- 112.9 \\
	& $\mathrm{CH_3 + HO_2 \rightarrow CH_4 + O_2}$\tablenotemark{b} & Two-body & 99.99 -- 10.66 -- 99.65 -- 11.45 & 0 -- 13.42 -- 34.21 -- 83.23 \\
	& $\mathrm{CH_3 + HNO \rightarrow NO + CH_4}$\tablenotemark{b} & Two-body & 12.44 -- 87.85 -- 10.47 & 5.66 -- 12.52 -- 22.15 \\
	& $\mathrm{CH_3CHO \rightarrow CH_4 + CO}$\tablenotemark{b} & Photo-dissociation & 12.96 -- 16.43 & 17.78 -- 19.51 \\
	& $\mathrm{H + CH_3 + M \rightarrow CH_4 + M}$\tablenotemark{b} & Three-body & 11.69 -- 99.48 -- 95.29 -- 96.96 & 65.75 -- 88.19 -- 103.57 -- 112.9 \\
	& & & & \\
	$\mathrm{N_2O}$ & $\mathrm{O(^1D) + N_2O \rightarrow NO + NO}$\tablenotemark{a} & Two-body & 62.1 -- 10.52 & 0 -- 20.39 \\
	& $\mathrm{O(^1D) + N_2O \rightarrow N_2 + O_2}$\tablenotemark{a} & Two-body & 37.9 -- 12.65 & 0 -- 17.78 \\
	& $\mathrm{N_2O \rightarrow N_2 + O(^1D)}$\tablenotemark{a} & Photo-dissociation & 15.65 -- 94.15 -- 92.39 -- 99.84 -- 91.73 & 14.31 -- 30.38 -- 48.92 -- 81.59 -- 112.9 \\
	& $\mathrm{NH_2 + NO_2 \rightarrow N_2O + H_2O}$\tablenotemark{b} & Two-body & 100 & 0 -- 30.38 \\
	& & Two-body & 100 -- 10.62 & 30.38 -- 37.19 \\
	& $\mathrm{N + NO_2 \rightarrow N_2O + O}$\tablenotemark{b} & Two-body & 42.8 -- 100 -- 99.99 & 36.19 -- 63.77 -- 112.9 \\
	& & & & \\
	\enddata
    \end{deluxetable*}
    \vspace{-0.5\baselineskip}
    \noindent\footnotesize
   \textbf{Notes.} $^{a}$ Destruction reaction; $^{b}$ Production reaction.
    \normalsize

\begin{figure*}[!ht]
	\centering
	\includegraphics[width=1\linewidth]{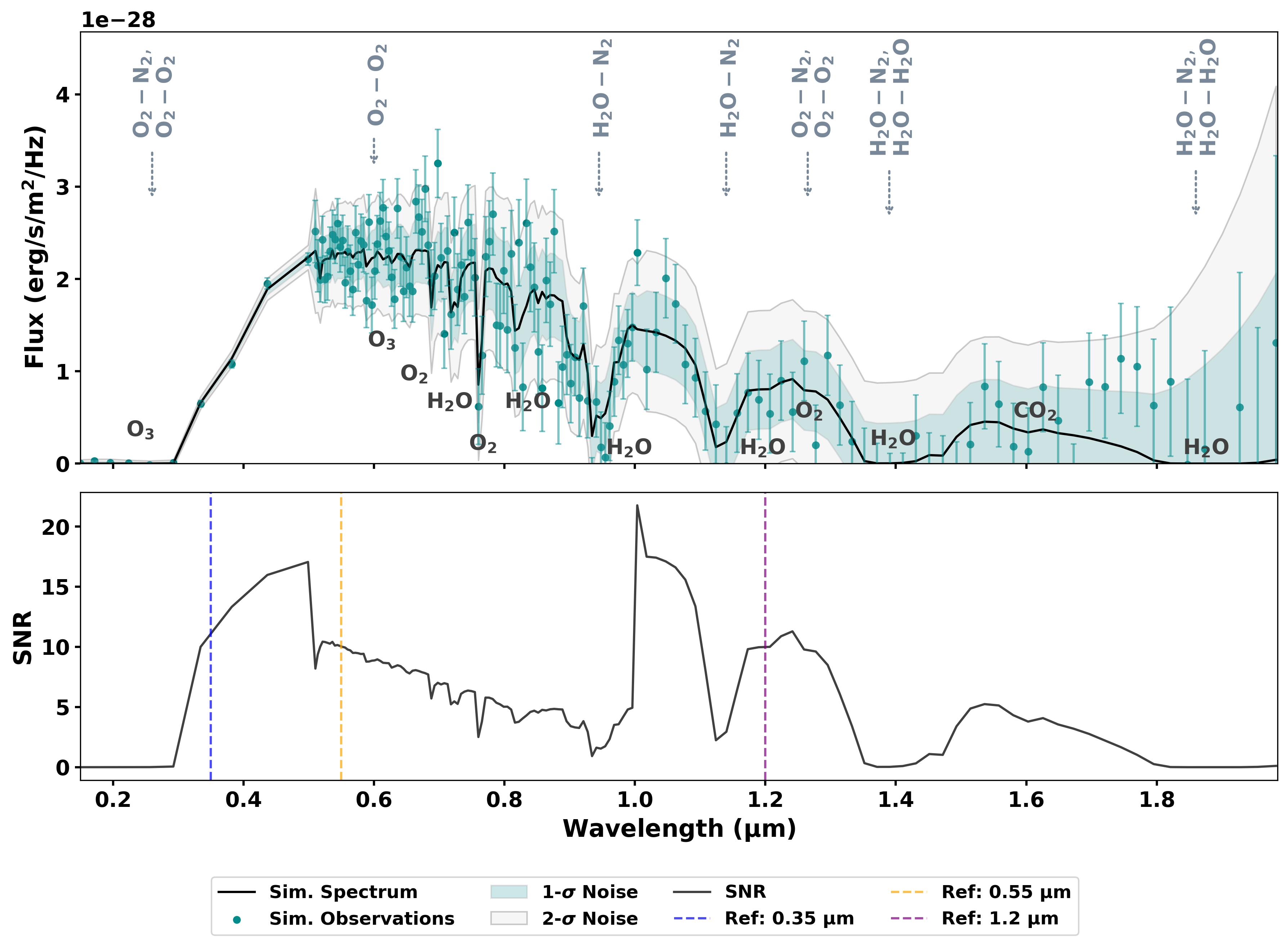}
	\caption{Top: simulated reflection spectra observation of modern Earth analog for the HWO mission concept. The simulated flux is considered for a planetary system at 10 pc distance. The wavelength is between 0.15 and 2 \textmu m. The $1 \sigma$ and $2 \sigma$ error bars are represented with cyan and light grey shaded area with solid edges respectively. The black curve is the spectrum from PSG. The cyan scatter plot represents the simulated observation points with corresponding uncertainties. The simulated reflection spectra is considered at a phase angle of $77.8^\circ$. Molecular features are labeled in black, and CIA features are marked in grey; Bottom: the associated SNR vs wavelength for the reflection spectra. The SNR is assumed to be 10 at three reference wavelengths: 0.35 \textmu m, 0.55 \textmu m and 1.2 \textmu m} 
	\label{fig:HWO_final_plot}
\end{figure*}

O$_3$ and H$_2$O are the primary species responsible for several spectral features observed at different wavelengths, including major dips at 0.3 \textmu m, 1.3--1.5 \textmu m, and 1.8--2 \textmu m in the HWO reflection spectra (see Figure \ref{fig:HWO_final_plot}). As expected, in the lower stratosphere, ozone absorbs incoming UV photons, causing the temperature of this layer to rise. In these layers, ozone is produced from O$_2$ through reactions involving a third body. Throughout the upper atmosphere (above 35 km), most O$_3$ destruction occurs via photochemical dissociation, while its production remains dominated by a three-body mediated reaction (contributing 100\%). In the lower atmosphere (0--20 km), its major destruction pathways involve photochemistry and two-body reactions.
\par
H$_2$O is the most prominent greenhouse gas in Earth's atmosphere. From the surface to the lifting condensation level (LCL, $\sim$2 km, where water begins to condense), the abundance of water vapor is maintained in balance between its production and destruction rates. However, as the rising air parcel cools, its capacity to hold water vapor decreases, leading to a reduction in its abundance. Beyond the LCL, condensation occurs, resulting in a further decrease in water vapor concentration up to the tropopause, beyond which transport is primarily mediated by diffusive mixing. This decline in H$_2$O abundance is further influenced by the two-body reaction involving O($^1$D) and OH, which contributes more than 90\% for altitudes up to $\sim$20 km. In the stratosphere and above, water vapor abundance remains relatively constant, as its initial concentration is primarily balanced by destruction processes involving O($^1$D) and UV photolysis (which produces H and OH radicals), as well as by its formation through the reaction of HNO$_3$ with OH. Beyond 80 km, H$_2$O abundance falls below 10$^{-7}$, rendering its chemistry insignificant.
\par
By fixing the CO$_2$ abundance at 400 ppm, as done by \citet{Tsai_2021}, we observe no significant change in its concentration across altitudes up to $\sim$80 km, beyond which it undergoes destruction due to photochemical dissociation and reactions with atomic nitrogen. The abundance of CO$_2$, primarily due to the major flux contribution from the lower layers of the atmosphere, remains largely unaffected by atmospheric chemistry. In these layers, CO$_2$ is primarily produced and destroyed through reactions with excited oxygen (O($^1$D)). The initial concentration of CO$_2$ at altitudes other than the surface remains consistent due to a near balance between its formation and destruction rates. However, destruction eventually outpaces production, which is compensated by the pre-existing OH and CO species, maintaining equilibrium.
\par
O$_2$ is a major photochemical product in Earth's modern atmosphere. Its abundance remains relatively constant up to altitudes of $\sim$85 km. Its major altitude-dependent formation and destruction reactions are detailed in Table \ref{tab:chem_rxns}. Regardless of its initial boundary conditions, O$_2$ remains well-mixed throughout the atmosphere up to about 80--85 km. 

In the troposphere, methane reacts with OH to produce H$_2$O, with reaction rates for this process slightly exceeding those contributing to methane formation, as shown in Table \ref{tab:chem_rxns}. Consequently, its abundance gradually decreases with altitude up to approximately 80 km, beyond which it declines more rapidly due to photochemical destruction.
\par
N$_2$O remains well-mixed in the troposphere, where the atmosphere is nearly devoid of N$_2$O sinks \citep{Tsai_2021}. The limited chemical reactions governing its production and destruction are detailed in Table \ref{tab:chem_rxns}. In the lower atmosphere (up to approximately 15 km), N$_2$O is primarily destroyed by O($^1$D). In the stratosphere, it is predominantly removed through photodissociation and is not effectively replenished by two-body reactions.

\begin{figure*}[!ht]
	\centering
	\includegraphics[width=\linewidth]{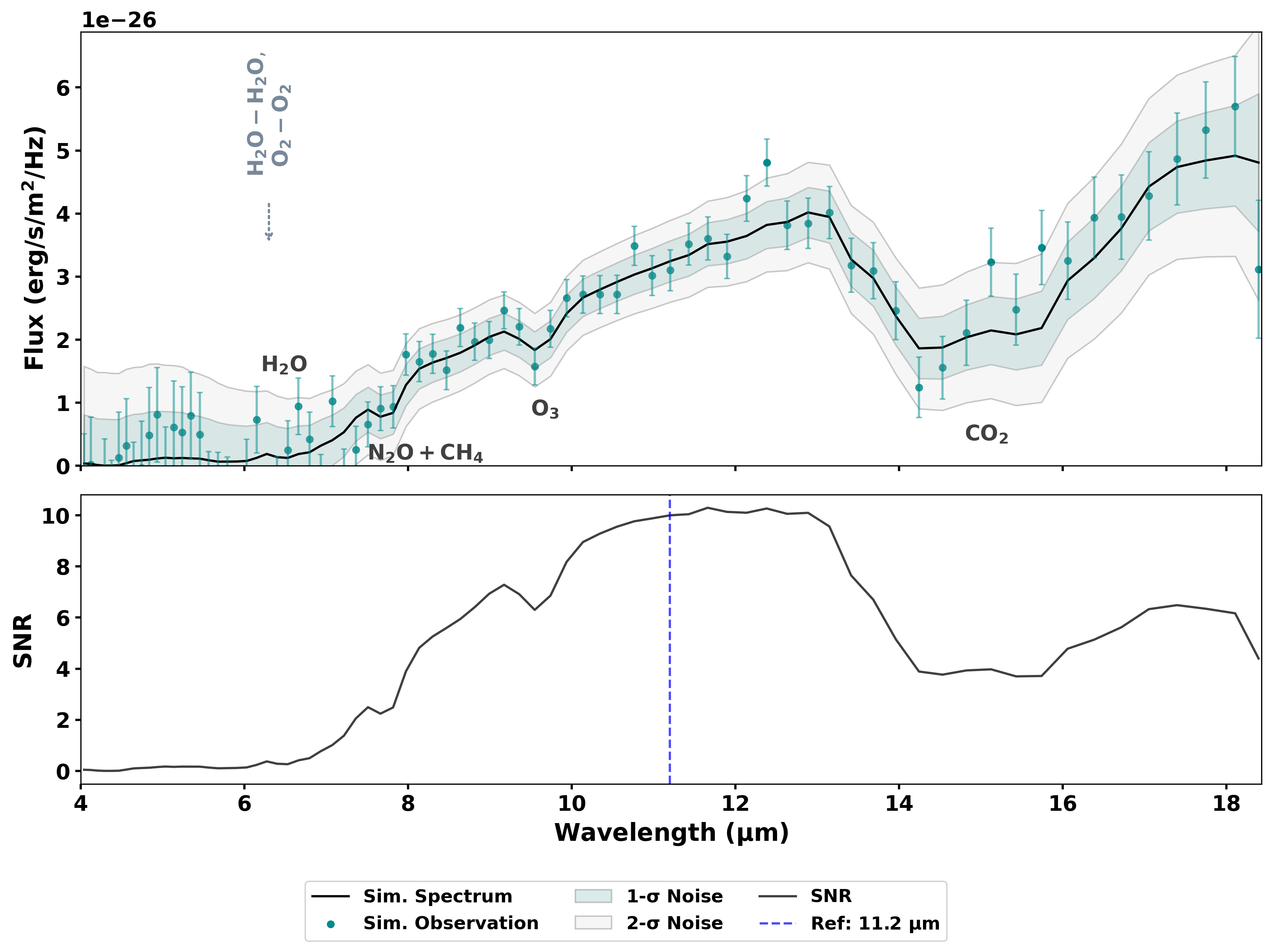}
	\caption{Top: simulated thermal emission spectra observation of modern Earth analog by the LIFE concept assuming a simplified noise. The simulated flux is considered for a planetary system at 10 pc distance. The wavelength is between 4 and 18.4 \textmu m. The $1 \sigma$ and $2 \sigma$ error bars are represented with cyan and light grey shaded area with solid edges respectively. The black curve is the spectrum from PSG fed into LIFEsim to simulate the observation (cyan scatter plot) with error bars. The simulated reflection spectrum is considered at a phase angle of $77.8^\circ$. Molecular features are labeled in black, and CIA features are marked in grey; Bottom: the associated SNR vs wavelength for the thermal emission spectra. The SNR is assumed to be 10 at the reference wavelength 11.2 \textmu m}
	\label{fig:LIFE_final_plot}
\end{figure*}

 \subsection{Reflection Spectra using \texttt{PSG}: Observables for HWO} \label{subsec:reflection}

In this section, we present the reflection spectra obtained from PSG, which is expected as a synthetic observation from HWO.  
As seen in Figure \ref{fig:HWO_final_plot}, the dips in the flux in the reflection spectra are primarily due to spectroscopically active molecules, namely $\mathrm{H_2O}$, $\mathrm{CO_2}$, $\mathrm{O_2}$, $\mathrm{CH_4}$, and $\mathrm{O_3}$. 

We have included Rayleigh scattering in our simulations. In Figure \ref{fig:HWO_final_plot}, which shows the flux, the sharp dip in the planet's flux as the wavelength decreases from 0.5 to 0.3 \textmu m corresponds to Rayleigh scattering due to nitrogen and oxygen. We define the spectral regions as ultraviolet (0.15–0.51 \textmu m), visible (0.51–1.0 \textmu m), and near-infrared (1.0–2.0 \textmu m). The sharp transitions at 0.51 \textmu m and 1.0 \textmu m shows detector-specific sensitivity regions and to reflect the varying resolving power relevant to HWO. In the reflection spectrum, O$_2$ exhibits distinct absorption features that can be used to infer its presence and concentration. Key features occur at wavelengths around 0.69 \textmu m and 0.76 \textmu m, corresponding to the O$_2$ B-band and A-band, respectively. The A-band at 0.76 \textmu m is particularly stronger than the B-band. Additionally, there is a very weak absorption feature at around 1.27 \textmu m, which is not distinguishable in Figure \ref{fig:HWO_final_plot} at low resolution, whereas features at other wavelengths remain prominent. Similarly, CH$_4$ has several spectral features, including relatively weak ones in the near-infrared (at 1.65 \textmu m), which can be detected at a resolution of 70.

$\mathrm{O_3}$ is photochemically produced from O$_2$ and, therefore, scales with atmospheric O$_2$ abundance. It exhibits strong spectral features across a range of wavelengths. Notable absorption occurs in the visible/near-infrared region (between 0.55 and 0.65 \textmu m) and in the ultraviolet ($\sim$0.26 \textmu m), both of which are evident in Figure \ref{fig:HWO_final_plot}. 
In the reflection spectrum shown in Figure \ref{fig:HWO_final_plot}, water vapor absorption features appear near 0.72 \textmu m, 0.82 \textmu m, 0.94 \textmu m, 1.13 \textmu m, 1.38 \textmu m, and 1.87 \textmu m. CO$_2$ exhibits weak features, particularly at 1.57 \textmu m, which stands out in the reflection spectrum (refer to Figure \ref{fig:HWO_final_plot}).

The CIA features identified in the spectra include $\mathrm{H_2O-N_2}$, $\mathrm{H_2O-H_2O}$, $\mathrm{O_2-N_2}$, and $\mathrm{O_2-O_2}$. 
The $\mathrm{H_2O-N_2}$ CIA is observed at 0.94 \textmu m, 1.12 \textmu m, 1.37 \textmu m, and 1.85 \textmu m, aligning closely with the major water vapor features in the spectrum. While the $\mathrm{H_2O-H_2O}$ CIA primarily contributes to the prominent water features at 1.38 \textmu m and 1.86 \textmu m, its impact is limited compared to the stronger contributions from $\mathrm{H_2O-N_2}$. Notably, the dips associated with $\mathrm{H_2O-N_2}$ are stronger than those from the $\mathrm{H_2O-H_2O}$ CIA; however, both remain relatively minor compared to the broad and deep absorption bands of water at these wavelengths, which result from fundamental vibrational transitions. The $\mathrm{O_2-N_2}$ CIA feature appears at 0.22 \textmu m in the UV region and enhances the weak $\mathrm{O_2}$ feature near 1.27 \textmu m, indicating the influence of nitrogen in the presence of oxygen. The $\mathrm{O_2-O_2}$ CIA contributes at similar wavelengths (0.22 \textmu m and $\sim$1.27 \textmu m), although its impact is less pronounced than that of $\mathrm{O_2-N_2}$. Additionally, $\mathrm{O_2-O_2}$ exhibits weaker features at 0.58 \textmu m and 0.63 \textmu m, where $\mathrm{O_3}$ is the dominant contributor. This analysis demonstrates that while collision-induced absorption plays a role, the dominant spectral features are still the intrinsic absorptions of $\mathrm{H_2O}$, $\mathrm{O_2}$, and $\mathrm{O_3}$.

We also present the expected noise levels in the spectra, with the corresponding SNR shown in the bottom plot of Figure \ref{fig:HWO_final_plot}, following the assumptions described in Section \ref{subsec: PSG noise SNR}. The SNR subplot closely follows the spectral pattern observed in the flux ratio plot. This correlation arises because the SNR is calculated as the ratio of the flux ratio signal at each wavelength to the 1$\sigma$ noise at that wavelength. Consequently, any dips in the flux ratio spectrum due to absorption features correspond to reductions in SNR at those wavelengths. The shaded regions in the main plot, representing 1$\sigma$ and 2$\sigma$ noise bounds, influence the depth of the SNR at specific wavelengths. Regions with broader noise bands indicate greater measurement uncertainty, leading to lower SNR values. This relationship between noise variations and SNR magnitude highlights how observational noise affects the detectability of spectral features, with the SNR plot emphasizing areas of higher and lower confidence in the signal based on the noise distribution.

The reflection spectra for all five Earth-like models, considering the boundary conditions discussed in Section \ref{subsec:test_models}, are shown in Figure \ref{fig:cases_spectra_HWO_grid_1} in Section \ref{subsec: reflctn}. A detailed analysis of the visible features and the governing chemistry is provided in Sections \ref{sec:apndx a} and \ref{subsec: reflctn}.

\subsection{Thermal Emission Spectra using \texttt{PSG} and \texttt{LIFEsim}: Observables for LIFE} \label{subsec:thermal_emission}

In this section, we present the thermal emission spectra (refer to Figure \ref{fig:LIFE_final_plot}) obtained from PSG, including the simulated synthetic observations expected from LIFE. As illustrated in the figure, the thermal emission spectra correspond to the presence of key spectrally active molecules such as $\mathrm{H_2O}$, $\mathrm{CO_2}$, $\mathrm{O_3}$, $\mathrm{CH_4}$, and $\mathrm{N_2O}$.

Additionally, we have calculated the expected noise levels in the spectra alongside the corresponding signal-to-noise ratio (SNR), both of which are illustrated in the bottom plot of Figure \ref{fig:LIFE_final_plot}.

$\mathrm{H_2O}$ shows strong absorption features in the mid-infrared around 6.3 \textmu m, which could be a signature of a moist atmosphere \citep{robinson2019earthexoplanet}. $\mathrm{CO_2}$ exhibits prominent absorption features at 4.3 \textmu m and around 15 \textmu m. $\mathrm{O_3}$, a product of photochemical reactions involving oxygen, shows a distinct absorption feature at 9.6 \textmu m. $\mathrm{CH_4}$ reveals its presence through spectral features in the mid-infrared around 7.7 \textmu m. Finally, $\mathrm{N_2O}$ displays a strong absorption feature around 7.7 \textmu m. $\mathrm{N_2O}$ also has a feature at $\sim$16.9 \textmu m, but it is not distinguishable as it is masked by overlapping spectral lines of $\mathrm{CO_2}$ and $\mathrm{H_2O}$ in the same region.

We observe that the only collision-induced absorption (CIA) features contributing to the spectra are $\mathrm{H_2O-H_2O}$ and $\mathrm{O_2-O_2}$. These features are located in a region dominated by a very strong $\mathrm{H_2O}$ absorption feature, making their contributions comparatively minor relative to the broad, deep, and continuous $\mathrm{H_2O}$ features. The $\mathrm{O_2-O_2}$ CIA is a small feature, contributing from 5.9 \textmu m to 7 \textmu m, with a peak at 6.44 \textmu m. In contrast, the $\mathrm{H_2O-H_2O}$ CIA has a much broader shape, with a base spanning from 5 \textmu m to 8 \textmu m and a peak around 6.17 \textmu m.

The noise levels for the thermal emission spectra, along with the corresponding SNR, are also presented in the bottom subplot of Figure \ref{fig:LIFE_final_plot}. These values were calculated following the description in Section \ref{subsec: LIFEsim noise SNR}. The thermal emission spectra are simulated over the wavelength range of 4 to 18.5 \textmu m with a constant resolution of R = 50, unlike HWO spectra. The SNR plot closely follows the features present in the spectrum. This means that as specific emission features appear in the spectrum, corresponding variations in the SNR are observed, with the SNR value for each feature shaped by the noise level around that wavelength. The consistent resolution allows for a clear representation of these features, indicating that the detectability of the signal is directly influenced by the spectral characteristics within this range.

The thermal emission spectra for all five Earth-like models, considering the varying boundary conditions discussed in Section \ref{subsec:test_models}, are shown in Figure \ref{fig:cases_spectra_LIFE_grid_1} in Section \ref{subsec: themlemisn}. An analysis of the observable features and the underlying chemistry is provided in Sections \ref{sec:apndx a} and \ref{subsec: themlemisn}.

\begin{figure*}[!ht]
	\centering
	
	\includegraphics[width=0.45\textwidth]{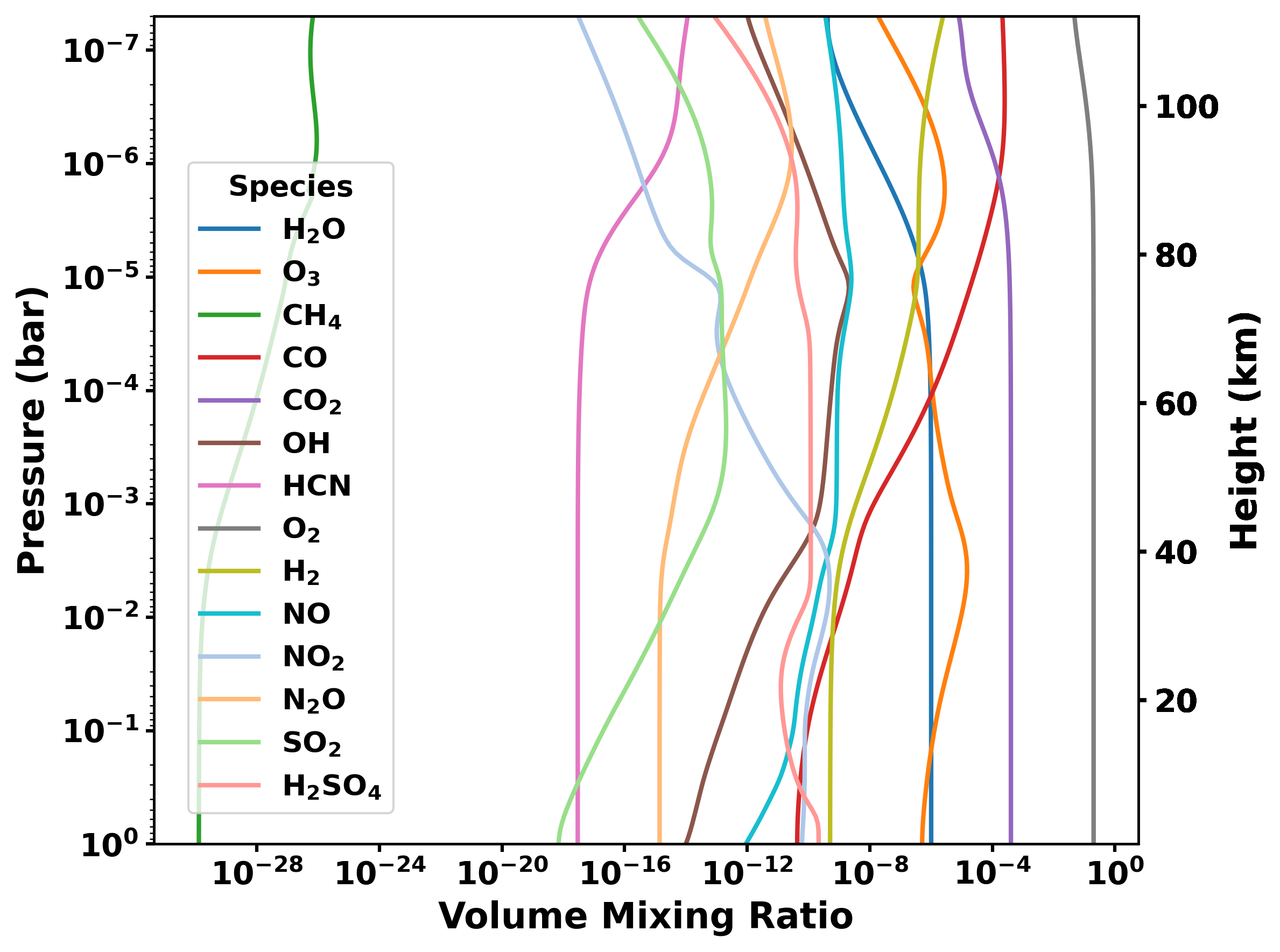}
	\hfill
	\includegraphics[width=0.45\textwidth]{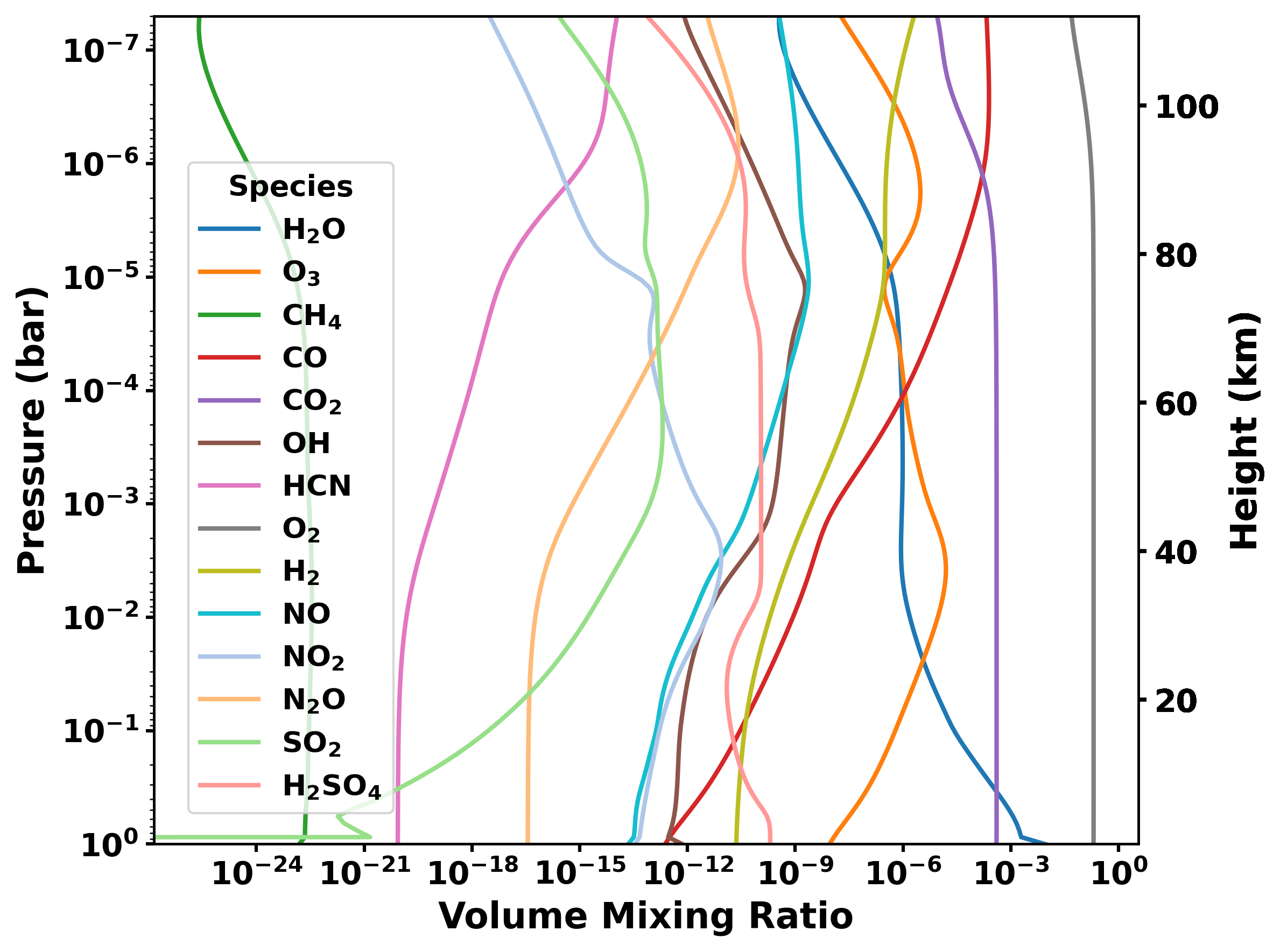}
	
	\vspace{0.5cm} 
	
	\includegraphics[width=0.45\textwidth]{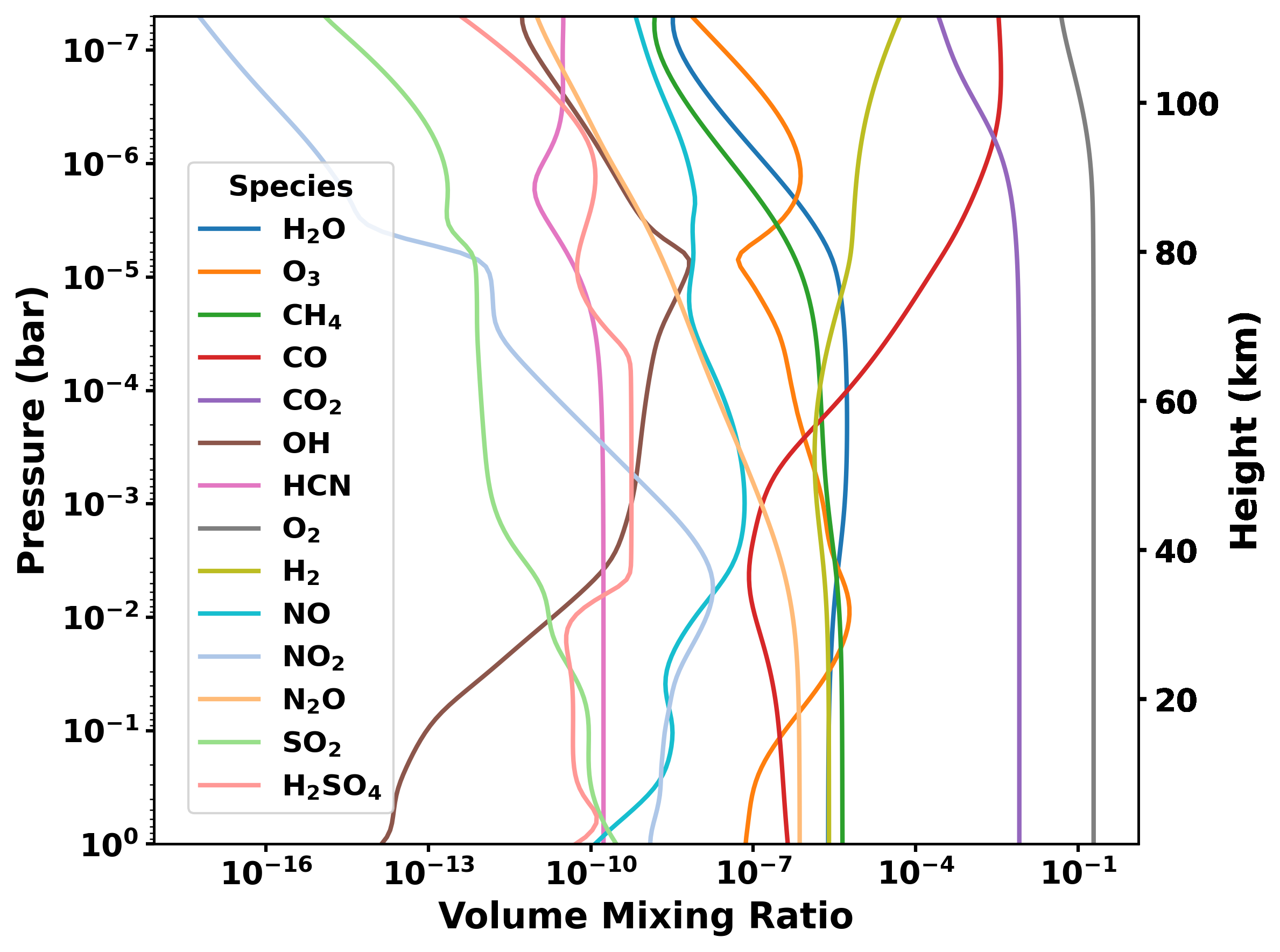}
	\hfill
	\includegraphics[width=0.45\textwidth]{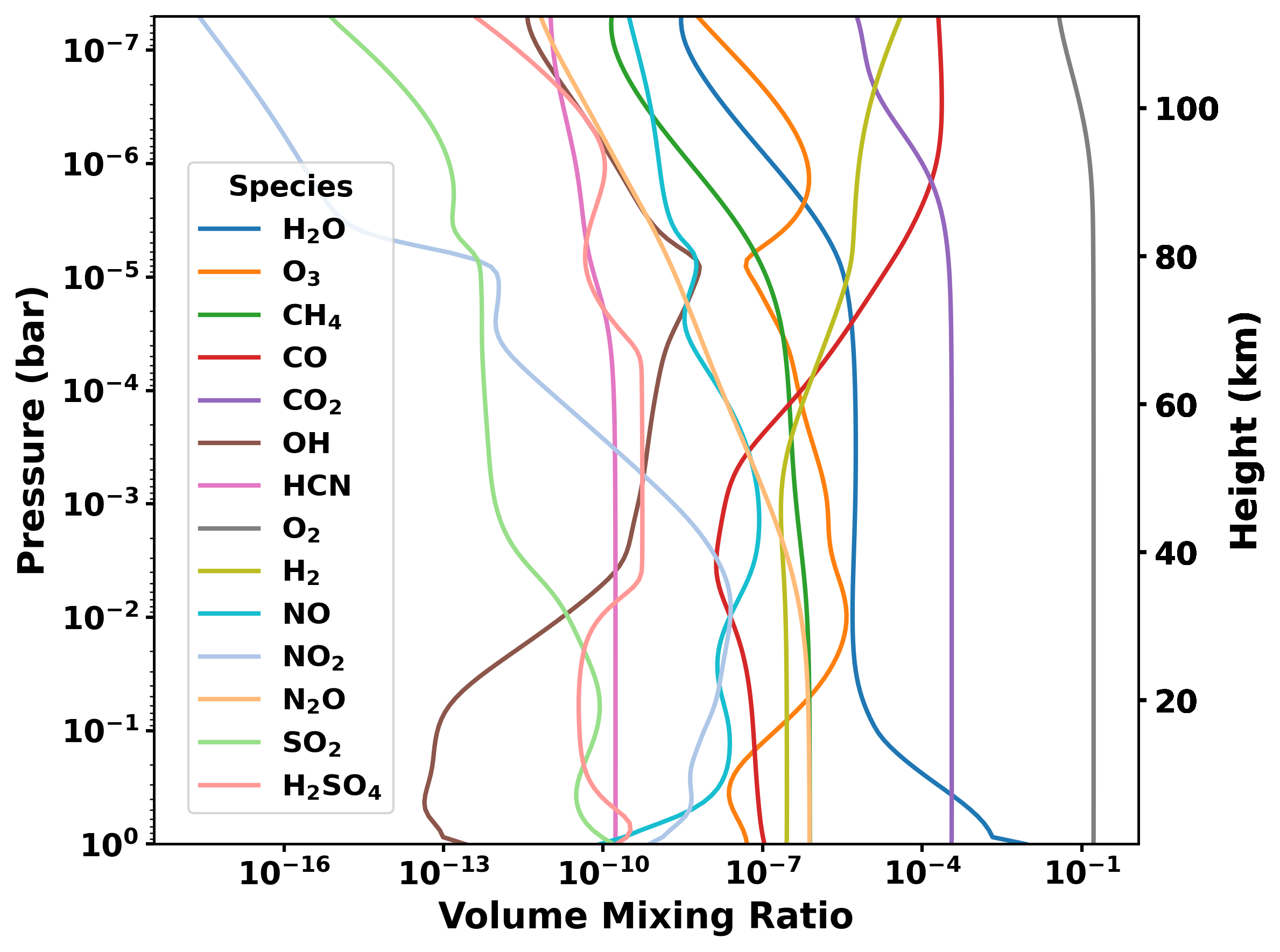}
	
	\vspace{0.5cm} 
	
	\includegraphics[width=0.45\textwidth]{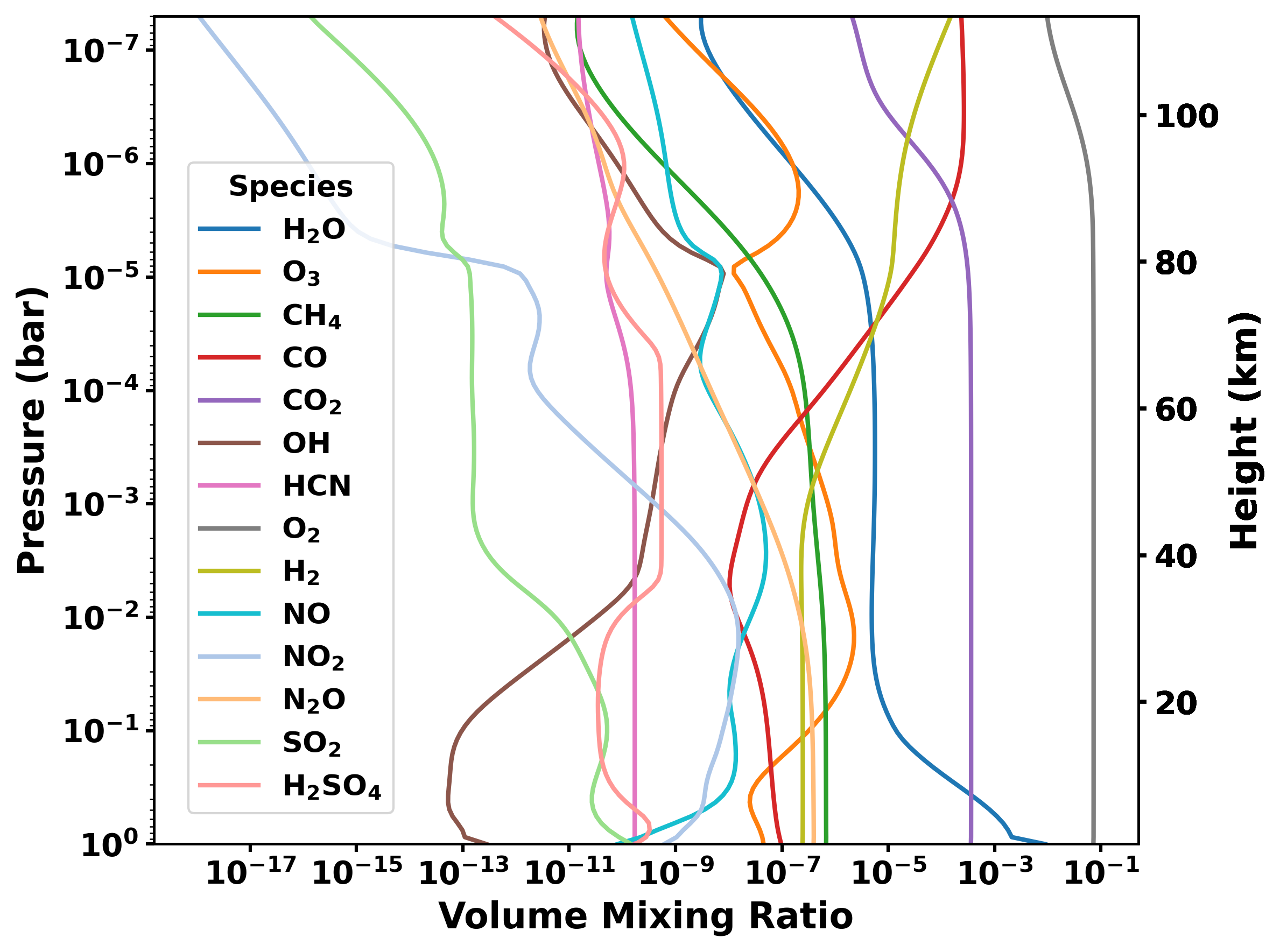}
	
	\caption{Comparison of mixing ratio profiles for Earth-like atmosphere by considering different boundary conditions (models M1 to M5) obtained from the kinetics model VULCAN, based on the temperature-sampled average TP profile. The profiles are organized into three rows: the top row displays models M1 and M2, the middle row shows models M3 and M4, and the bottom row features model M5. Each subplot illustrates the mixing ratios for specific atmospheric molecules: $\mathrm{H_2O}$, $\mathrm{O_3}$, $\mathrm{CH_4}$, $\mathrm{CO}$, $\mathrm{CO_2}$, $\mathrm{OH}$, $\mathrm{HCN}$, $\mathrm{O_2}$, $\mathrm{H_2}$, $\mathrm{NO}$, $\mathrm{NO_2}$, $\mathrm{N_2O}$, $\mathrm{SO_2}$, $\mathrm{H_2SO_4}$, providing a comprehensive view of molecular distribution across the different models.}
	\label{fig:cases_grid}
\end{figure*}

\subsection{Atmospheric Chemistry for Earth-like Models with Varying Boundary Conditions}\label{sec:earth_model_descrptn} \label{sec:apndx a}

In this section, 
we present the Earth-like model simulations (as described in Section \ref{subsec:test_models}) considering variations in the bottom boundary conditions that reflect biological and geological processes. We discuss their chemistry and in the following section, assess how they influence the atmospheric spectra. A summary of the models can be found in Table \ref{tab:atmospheric-models}. 

The mixing ratio profiles for all the models are shown in Figure \ref{fig:cases_grid}. In the figure, we notice that for models where we have a bottom boundary, H$_2$O abundance decreases up to $\sim$20 km when we fix its surface mixing ratio. Beyond which H$_2$O VMR is very well mixed until around 80 km and then further decreases due to photochemical reactions. However, when bottom boundary conditions are not applied and H$_2$O mixing ratio is not fixed at the surface, it remains well-mixed from the surface level without showing any significant deviations below 20 km. For CH$_4$, the VMR gets significantly affected by the boundary conditions throughout the atmosphere. In the absence of bottom boundary conditions, CH$_4$ shows major VMR change $\sim$10$^{-22}$ -- 10$^{-30}$ from 10$^{-6}$ (when bottom boundary remains present), whereas CO decreases by four orders of magnitude, as expected, due to the absence of its surface flux. CO$_2$, however, remains abundant at 400 ppm even when there is neither its source nor its sink, along with no fixed surface abundance. Nitrogen-bearing species such as N$_2$O in the absence of tropospheric sinks keep on building up in the atmosphere due to surface sources up to a VMR of $\sim$10$^{-6}$. However, its VMR falls due to photodissociation in the stratosphere and beyond. In the absence of N$_2$O bottom boundary, its VMR does not go above $\sim$10$^{-12}$. In all these models, O$_2$ remains highly abundant.

\par
\begin{figure*}[!ht]
	\centering
	\includegraphics[width=\linewidth]{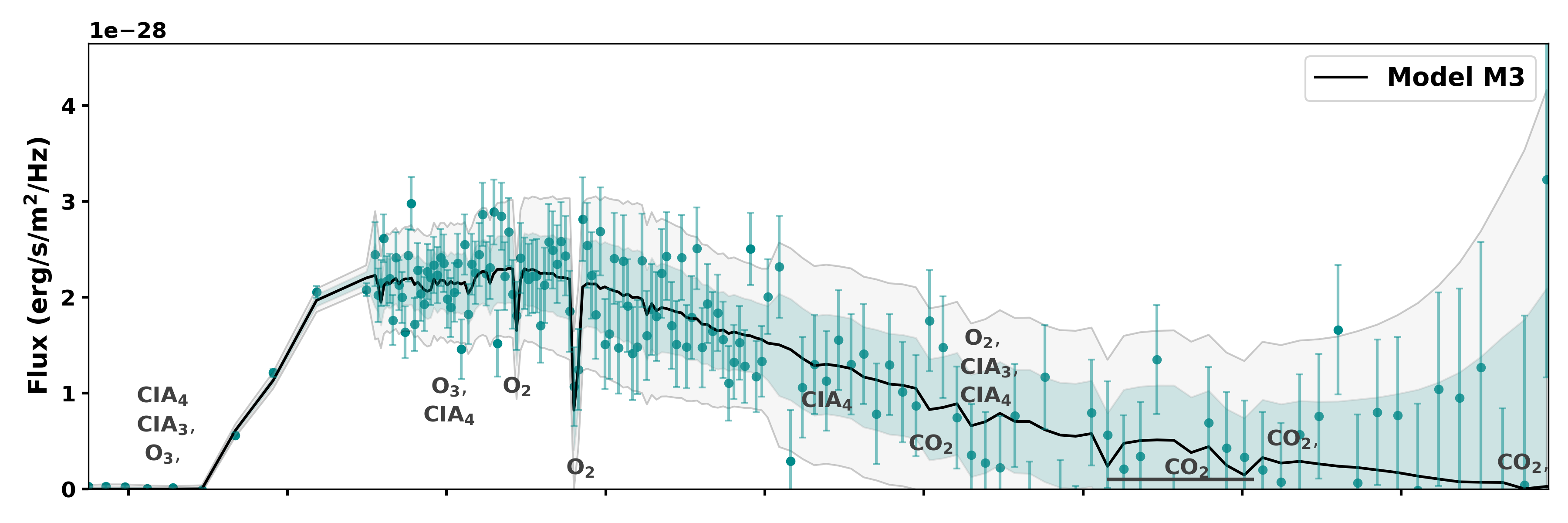}
	\vspace{0.3cm}
	
	\includegraphics[width=\linewidth]{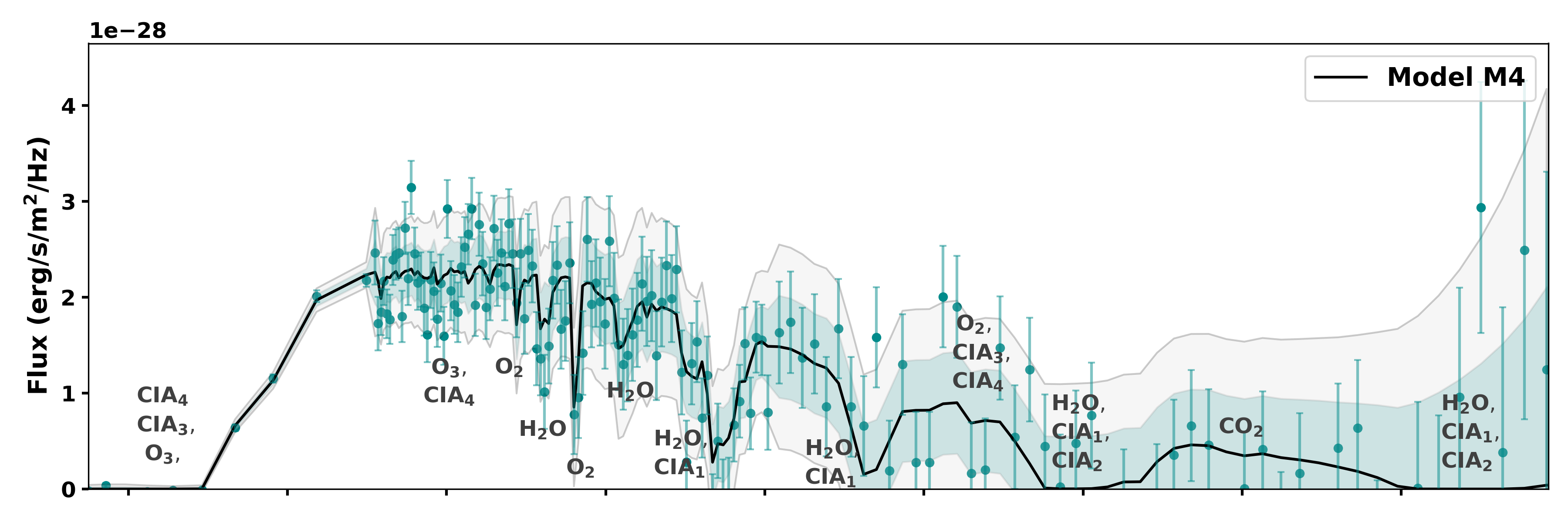}
	\vspace{0.3cm}
	
	\includegraphics[width=\linewidth]{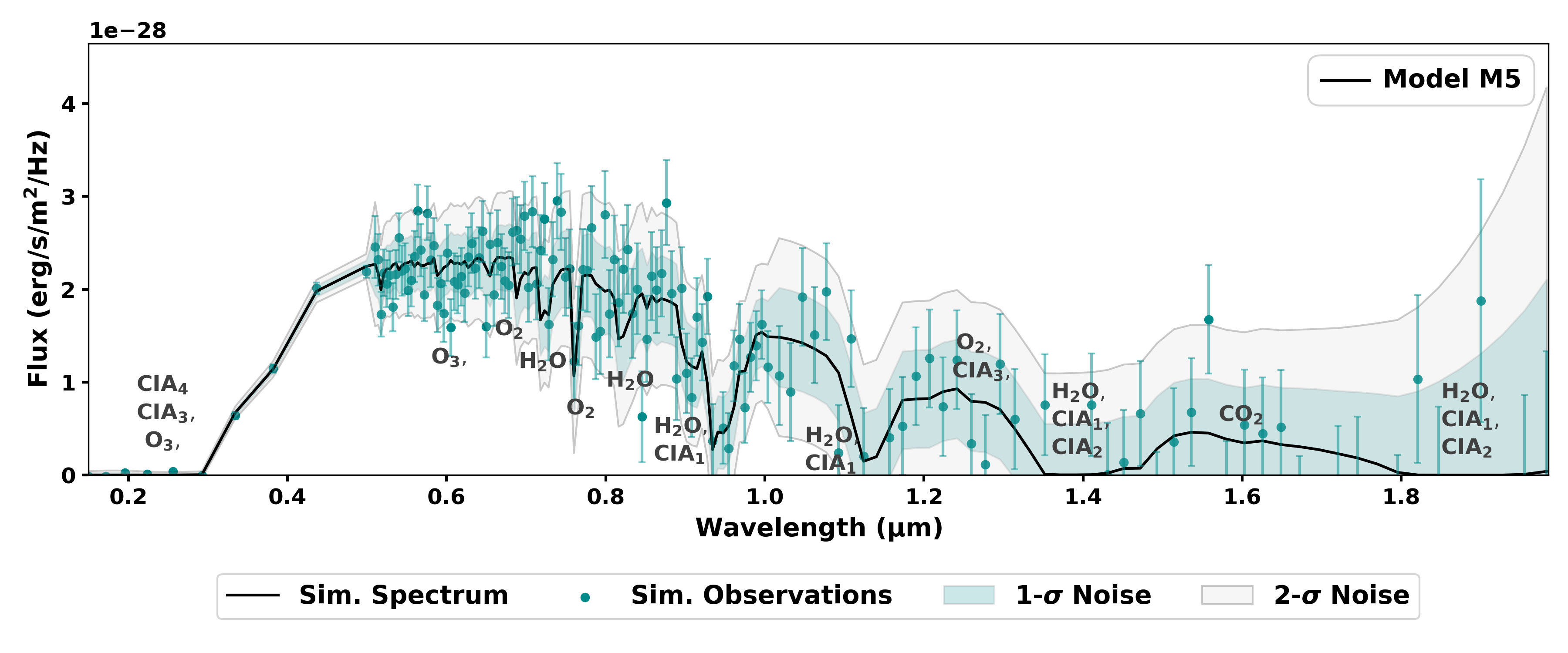}
	\caption{Simulated reflection spectra observation of Earth analog by considering different boundary conditions in the models M1 to M5 (top to bottom) for the HWO mission concept. The simulated fluxes are considered for a planetary system at 10 pc distance. The wavelength is between 0.15 and 2 \textmu m. The $1 \sigma$ and $2 \sigma$ error bars are represented with cyan and light grey shaded area with solid edges respectively. The black curve is the spectrum from PSG. The cyan scatter plot represents the simulated observation points with corresponding uncertainties. The simulated reflection spectrum is considered at a phase angle of $77.8^\circ$. Here, $\mathrm{CIA_1}$, $\mathrm{CIA_2}$, $\mathrm{CIA_3}$, and $\mathrm{CIA_4}$ refer to $\mathrm{H_2O-N_2}$, $\mathrm{H_2O-H_2O}$, $\mathrm{O_2-N_2}$, and $\mathrm{O_2-O_2}$ CIAs respectively.}
	\label{fig:cases_spectra_HWO_grid_1}
\end{figure*}

In HWO spectra for M1 and M3 as shown in Figure \ref{fig:cases_spectra_HWO_grid_1} in Section \ref{subsec: reflctn}, H$_2$O is seen to be absent at $\sim$0.9 \textmu m, 1.1 \textmu m, and 1.4 \textmu m compared to the rest of the cases. The reactions responsible for H$_2$O consumption are as follows:
(i) $\mathrm{SO}_3\mathrm{ + H}_2\mathrm{O} \rightarrow \mathrm{H}_2\mathrm{SO}_4$ contributing 59\% -- 94\% for upto 4.5 km and $<$29\% beyond, and
(ii) $\mathrm{O(^1D) + H}_2\mathrm{O} \rightarrow \mathrm{OH + OH}$ contributing upto 42\% within 4.5 km, till 100\% for upto $\sim$21 km. In addition to the chemical reactions that continuously produce H$_2$O, the supply of a fixed H$_2$O abundance at the surface (see Table \ref{tab:atmospheric-models}) is necessitated by the appearance of H$_2$O feature in the HWO spectra. When the surface abundance of H$_2$O is fixed, the aforementioned chemical reactions will be unable to deplete the H$_2$O abundance near the surface as they do in the upper atmosphere (above $\sim$20 km). At $\sim$1.55 \textmu m and $\sim$1.96 \textmu m in models M1 and M3, prominent CO$_2$ features are present because of the absence of H$_2$O (more prominent in M3) and contribute similar to Table \ref{tab:chem_rxns}. For M3, upto 6.6 km, a reaction involving singlet oxygen contributes to CO$_2$ formation from 82 -- 91\% ($\sim$100\% for M1), while the same reaction destroys CO$_2$ ($\sim$100\%) and the net CO$_2$ abundance does not change from the initial atmospheric abundance due to this dominant reaction.
This is in contrast to the atmosphere having fixed boundary conditions (Table \ref{tab:chem_rxns}) where the formation reaction of CO$_2$ at lower atmosphere is dominated by the reaction involving OH and CO. In M3, the reactants OH and CO contribute upto a maximum of 18\% near the surface (upto $\sim$5 km). This reaction plays an important role as the second dominating reaction when the major reaction involving O($^1$D) almost nullifies the production of CO$_2$ with its destruction. So, the abundance of CO from the surface emission also impacts the abundance of CO$_2$ at any particular atmospheric layer. In the absence of boundary (M1 and M2), the lack of CO source could directly deplete the CO$_2$ abundance while in the presence of boundary with CO$_2$ and O$_2$ fluxes (M4 and M5), CO$_2$ surface sink could decrease the CO$_2$ abundance. However, in M3 there is a continuous CO flux as well as no CO$_2$ surface sink. Consequently, the abundance of CO$_2$ in M3 becomes slightly higher than the rest of the models. Above 90 km, the destruction reaction involving Nitrogen and the production reaction involving O($^1$D) and CO starts to exhibit major contribution, surpassing the contribution of the reaction (involving O($^1$D) and CO$_2$) which dominates the lower atmosphere.
\par
For LIFE spectra in M1 and M2 (no boundary) as shown in Figure \ref{fig:cases_spectra_LIFE_grid_1} in Section \ref{subsec: themlemisn}, CH$_4$ and N$_2$O at 7.9 \textmu m remain absent due to absence of their continuous surface flux. Whatever minor amount is produced gets consumed by the reactions as shown in Table \ref{tab:chem_rxns}. For N$_2$O, the first two destruction reactions of Table \ref{tab:chem_rxns} for M1 and M2 remain 62\% and 38\% respectively across all layers upto 15 km, beyond which the destruction contribution starts to fall. In the absence of continuous surface emission flux, N$_2$O abundance diminishes to $\sim$10$^{-16}$, which is orders of magnitude below detectable thresholds of LIFE.
Similar to N$_2$O, the detectability of CH$_4$ also strongly depends on its surface emission flux. In the absence of continuous surface emission of CH$_4$ (M1 and M2), the initial abundance of CH$_4$ is majorly chemically destroyed by OH upto an altitude of 24.7 km, contributing $\sim$50 -- 98\% ($\sim$97 -- 100\% for M2), followed by the reaction:
$\mathrm{O(^1D) + CH}_4 \rightarrow \mathrm{OH + CH}_3$ contributing upto 47\%. This reaction dominates above 24.7 km upto $\sim$73 km, beyond which the CH$_4$ destruction is dominated by photodissociation to $^1$CH$_2$ (CH$_2$ singlet). In the LIFE spectra the presence of stronger CO$_2$ feature at $\sim$15 \textmu m (M3), as described previously, can be attributed to continuous CO flux and absence of CO$_2$ surface sink.
\subsection{Simulated Reflection and Thermal Emission Spectra for the Earth-like Models} \label{sec:apndx c}

In this section, we present the simulated reflection and thermal emission spectra for all five models with varying boundary conditions (see Table \ref{tab:atmospheric-models} in Section \ref{sec:apndx a}). The reflection spectra are provided in Section \ref{subsec: reflctn}, and the thermal emission spectra are shown in Section \ref{subsec: themlemisn}. We then discuss the molecular features that are evident in each spectrum. Furthermore, we consider how changes in boundary conditions impact the strength and presence of these spectral features.

\onecolumngrid

\begin{figure*}[!hb]
	\centering
	\includegraphics[width=\linewidth]{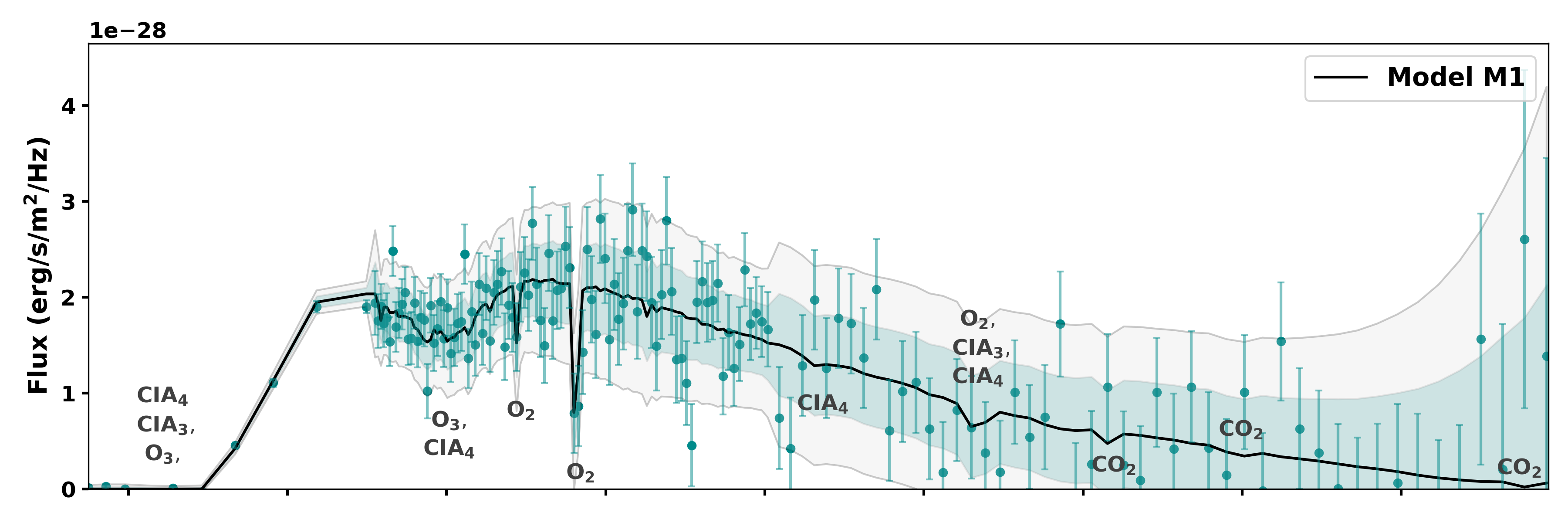}
	\vspace{0.3cm}
	
	\includegraphics[width=\linewidth]{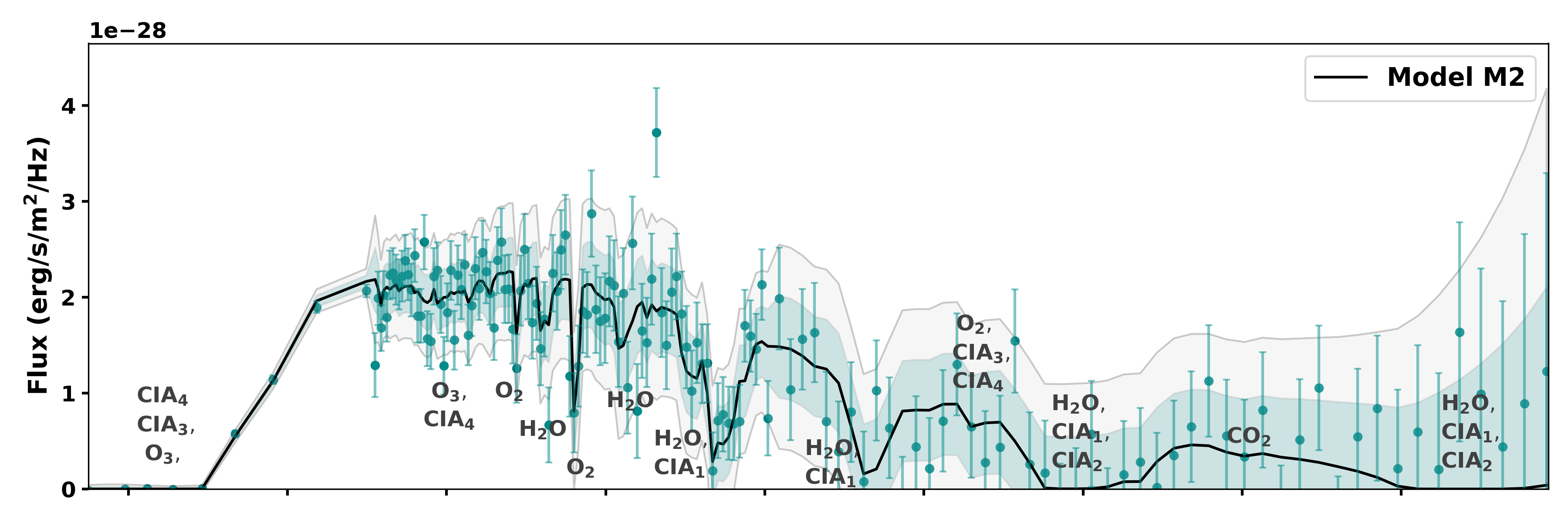}
	
\end{figure*}

\clearpage
\twocolumngrid

\subsubsection{Reflection Spectra} \label{subsec: reflctn}

For models M1 and M3 in the Figure \ref{fig:cases_spectra_HWO_grid_1}, $\mathrm{H_2O}$ absorption features are absent around $\sim 0.9 \, \mu \mathrm{m}$, $1.1 \, \mu \mathrm{m}$, and $1.4 \, \mu \mathrm{m}$ compared to the other models. This is due to the lower $\mathrm{H_2O}$ abundance, as shown in Figure \ref{fig:cases_grid}. The absence of detectable $\mathrm{H_2O}$ in the spectrum is linked to the lack of a fixed $\mathrm{H_2O}$ surface abundance and the reactions responsible for $\mathrm{H_2O}$ consumption, as detailed in Section \ref{sec:apndx a}. Apart from this, other spectral features are still evident in these two models.

At $\sim 1.55 \, \mu \mathrm{m}$ and $\sim 1.96 \, \mu \mathrm{m}$, prominent $\mathrm{CO_2}$ features are observed, particularly in the absence of $\mathrm{H_2O}$, and are especially strong in model M3. In model M3, $\mathrm{CH_4}$ is also clearly visible, unlike in model M1, because model M3 includes $\mathrm{CH_4}$ under its default boundary conditions. In contrast, in model M2, even without default boundary conditions, the key features from the main model are still prominent. This is because molecules without defined fluxes or deposition velocities in the boundary file do not show features in this region (except for $\mathrm{CH_4}$). $\mathrm{CH_4}$ is not present in the boundary for model M2, and therefore, does not appear in its spectrum. However, $\mathrm{O_2}$ and $\mathrm{O_3}$ behave differently. Their abundances are high across all models, as shown in Figure \ref{fig:cases_spectra_HWO_grid_1}, so their features are consistently visible.  In the last two models, where we introduced a $\mathrm{CO_2}$ boundary and varied the oxygen boundary while keeping other molecules fixed, we confirm that the abundances of $\mathrm{O_2}$, $\mathrm{O_3}$, and $\mathrm{CO_2}$ are not significantly affected by these changes. As a result, their spectral features remain prominent in these models.

\onecolumngrid
\FloatBarrier
\begin{figure*}[!hb]
	\centering
	\includegraphics[width=\linewidth]{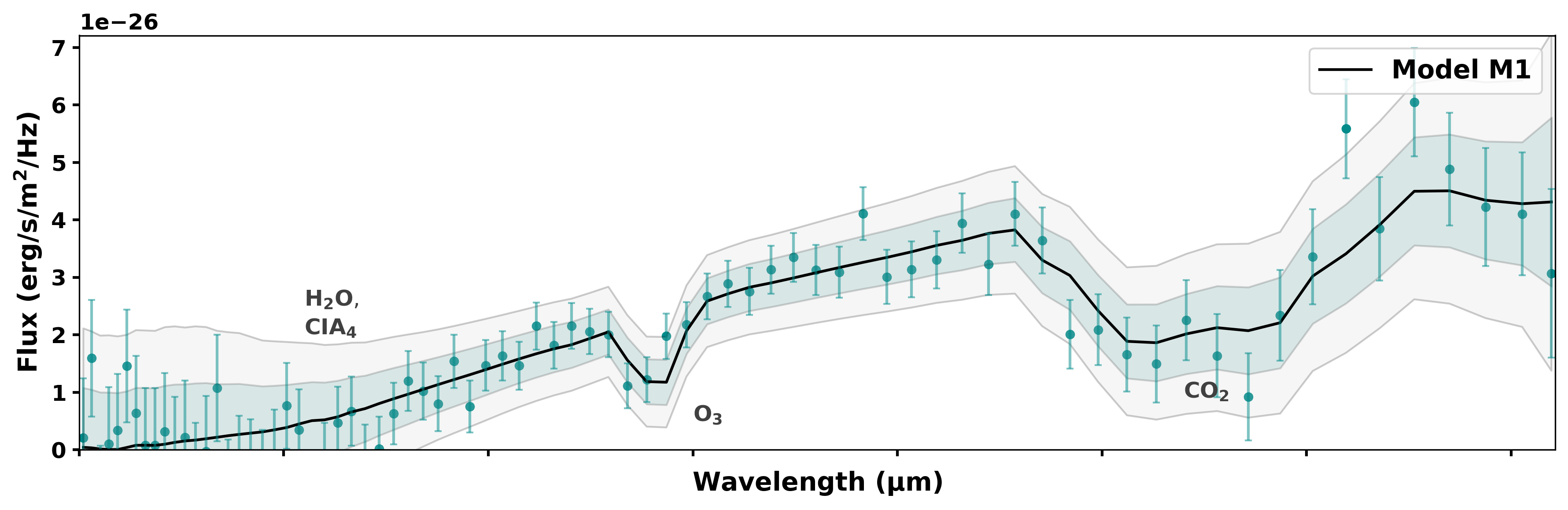}
	\vspace{0.3cm}
	
	\includegraphics[width=\linewidth]{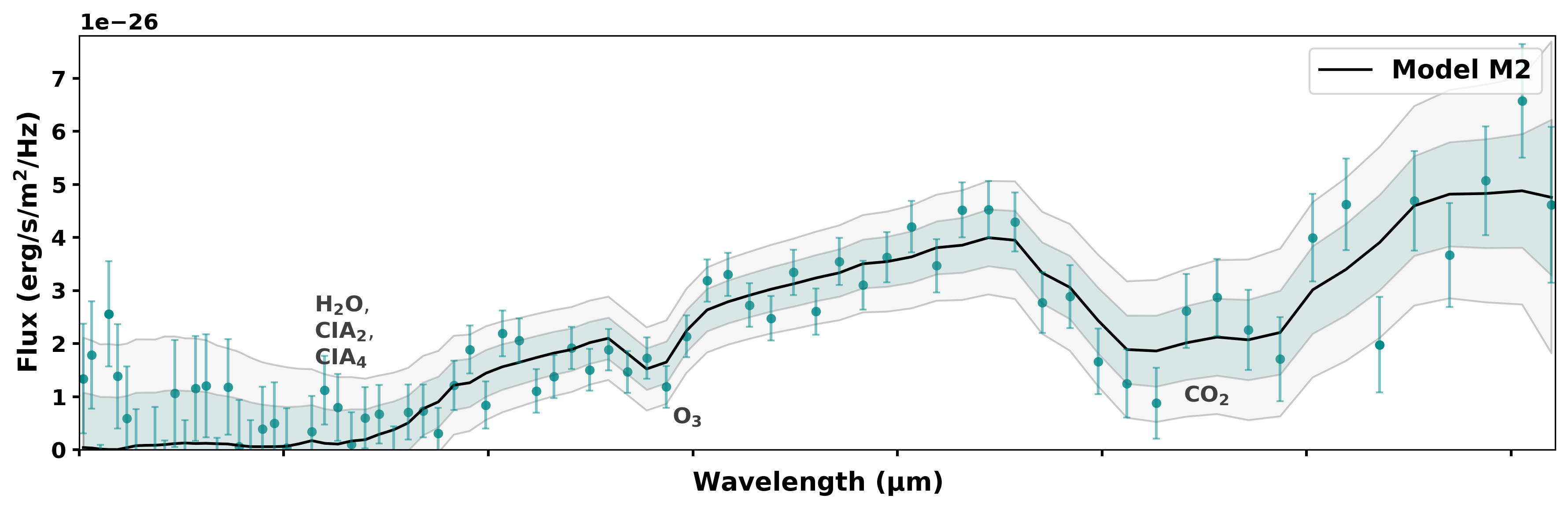}
\end{figure*}
\begin{figure*}[!ht]
	\centering
	\includegraphics[width=\linewidth]{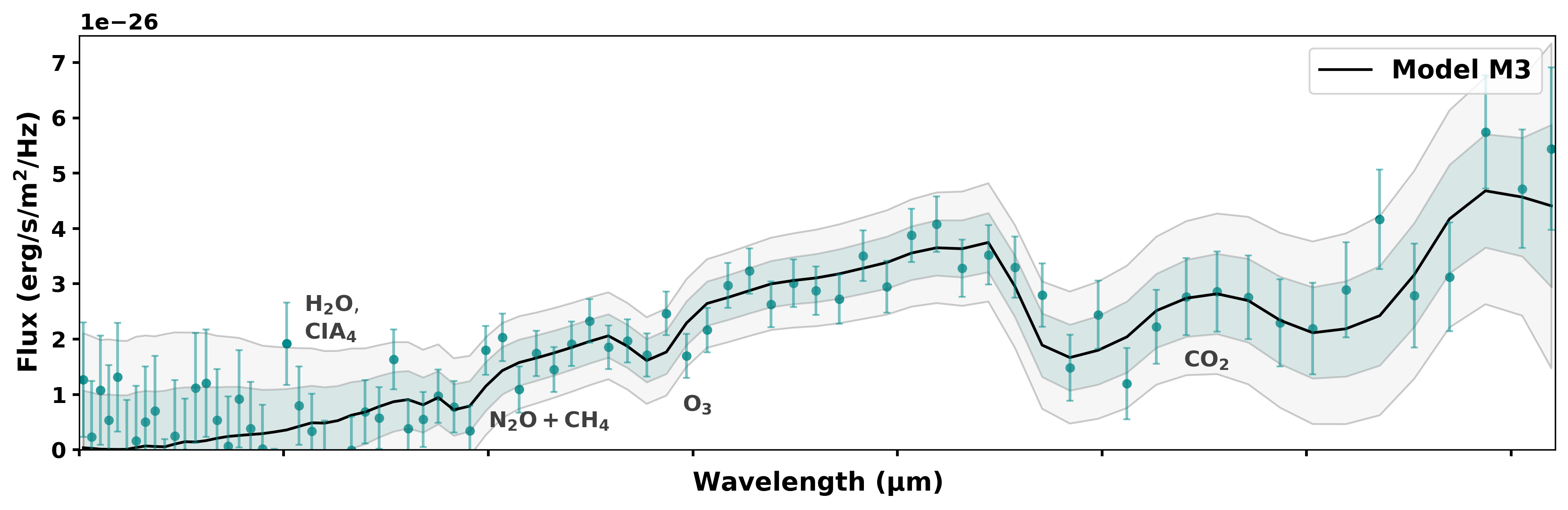}
	\vspace{0.3cm}
	
	\includegraphics[width=\linewidth]{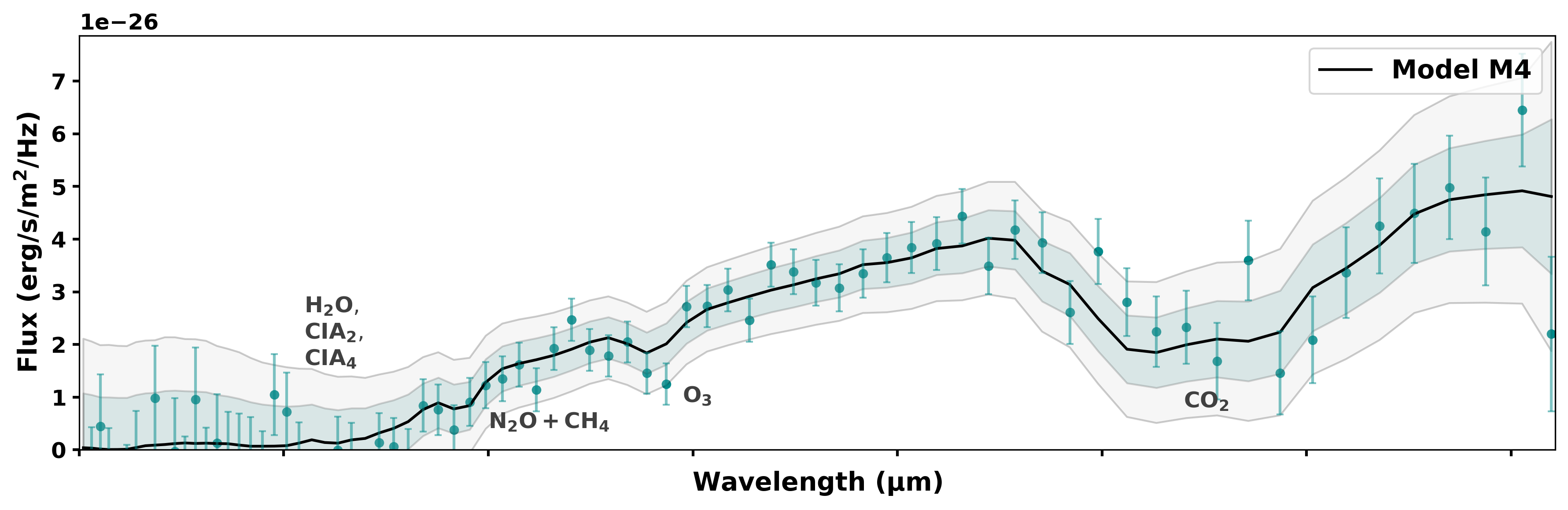}
	\vspace{0.3cm}
	
	\includegraphics[width=\linewidth]{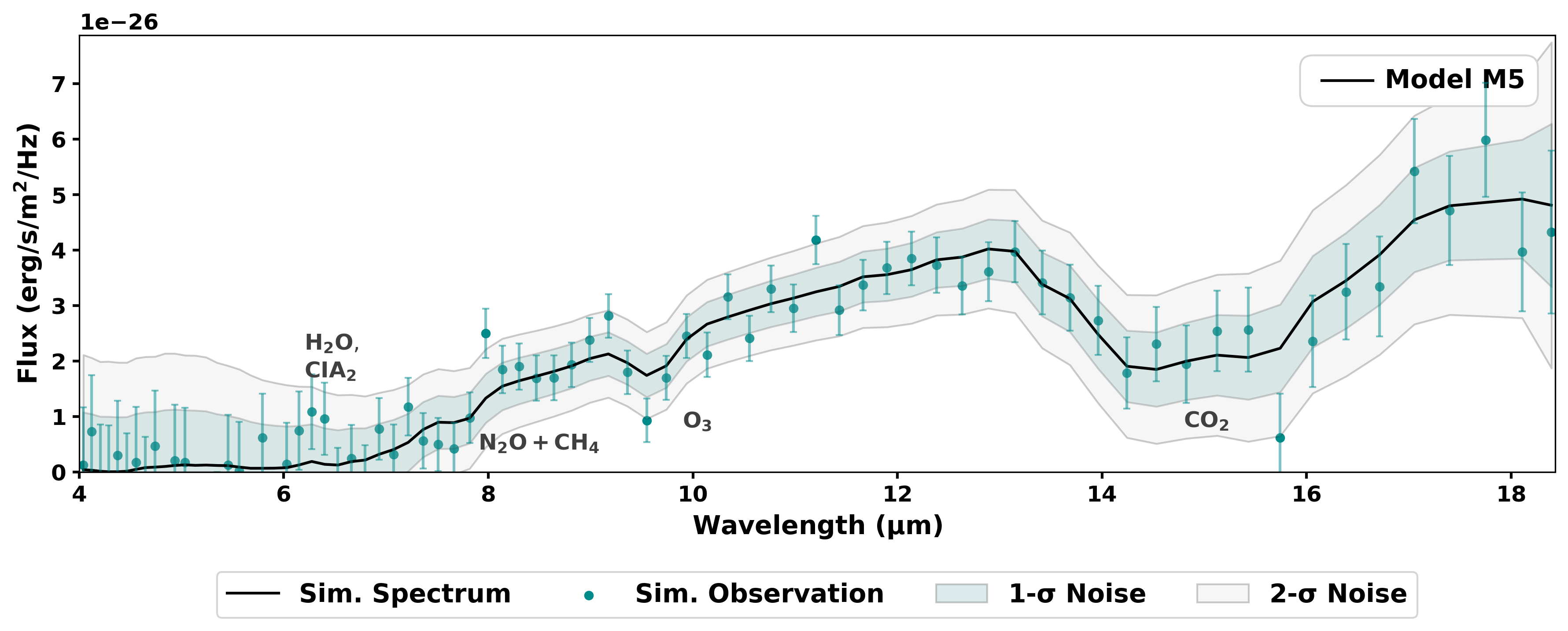}
	
	\caption{Simulated thermal emission observation of Earth analog by considering different boundary conditions in the models M1 to M5 (top to bottom) for LIFE. The simulated fluxes are considered for a planetary system at 10 pc distance. The wavelength is between 4 and 18.5 \textmu m. The $1 \sigma$ and $2 \sigma$ error bars are represented with cyan and light grey shaded area with solid edges respectively. The black curve is the spectrum from PSG fed into LIFEsim. The cyan scatter plot represents the simulated observation points with corresponding uncertainties obtained from LIFEsim. The simulated thermal emission spectrum is considered at a phase angle of $77.8^\circ$. As like HWO spectra, here, $\mathrm{CIA_2}$ and $\mathrm{CIA_4}$ refer to  $\mathrm{H_2O-H_2O}$ and $\mathrm{O_2-O_2}$ CIAs respectively.}
	\label{fig:cases_spectra_LIFE_grid_1}
\end{figure*}

\twocolumngrid
\FloatBarrier

We compare the CIA contribution in these models with those identified in our base spectra (see Figure \ref{fig:HWO_final_plot}), which include $\mathrm{H_2O-N_2}$, $\mathrm{H_2O-H_2O}$, $\mathrm{O_2-N_2}$, and $\mathrm{O_2-O_2}$. Notably, $\mathrm{H_2O-N_2}$ and $\mathrm{H_2O-H_2O}$ CIA are absent in models M1 and M3, attributed to the very low abundance of H$_2$O in these simulations. However, the CIA contributions from $\mathrm{O_2-N_2}$ and $\mathrm{O_2-O_2}$ are present in all the spectra across their respective wavelengths, as discussed in Section \ref{subsec:reflection}. This consistent presence is due to the robustness of O$_2$ detection, which is not significantly affected by boundary conditions. Additionally, we observe a faint feature of $\mathrm{O_2-O_2}$ CIA around 1.08 \textmu m in models M1 and M3, which is otherwise obscured in the base model and other configurations due to overlapping water bands. Interestingly, $\mathrm{O_2-O_2}$ CIA is nearly absent in model M5. We note that O$_2$ VMR in this model is slightly below 0.1 (see Figure \ref{fig:cases_spectra_HWO_grid_1}) as compared to other models suggesting that the VMR threshold for O$_2$ to contribute to this particular CIA feature is about 0.1.

\subsubsection{Thermal Emission Spectra} \label{subsec: themlemisn}

For models M1 and M2 in Figure \ref{fig:cases_spectra_LIFE_grid_1}, the CH$_4$ and N$_2$O features around 7.9 \textmu m are absent, as their abundances (shown in Figure \ref{fig:cases_grid}) are too low to be detectable in the spectra. Any minor production is quickly consumed. Additionally, in model M1, the H$_2$O feature is too faint to be visible due to the lack of a specified fixed surface boundary for H$_2$O, resulting in very low production. Similarly, in model M3, the absence of a fixed bottom boundary for H$_2$O leads to its negligible presence in the spectra. However, in model M3, CH$_4$ and N$_2$O are observable in the spectra due to the default boundary conditions that allow for sufficient production of these molecules. In models M4 and M5, despite varying O$_2$ boundary conditions, all expected spectral features remain visible, confirming that O$_2$ and O$_3$ are not significantly affected by these boundary conditions. The strong CO$_2$ features present in all models are due to a balance between its production and destruction processes, keeping its levels consistent.

As discussed in Section \ref{subsec:thermal_emission}, the only CIA features contributing to the thermal emission spectra are $\mathrm{H_2O-H_2O}$ and $\mathrm{O_2-O_2}$. Among these, the contribution from $\mathrm{H_2O-H_2O}$ is absent in models M1 and M3, as anticipated due to the very low abundance of H$_2$O in these configurations. In the other models, however, $\mathrm{H_2O-H_2O}$ contributes. The $\mathrm{O_2-O_2}$ feature is present in all models, due to the high abundance of O$_2$. However, as noted in Section \ref{subsec: reflctn}, this CIA feature does not contribute in model M5, suggesting that the abundance of O$_2$ must exceed 0.1 for $\mathrm{O_2-O_2}$ CIA to play a role.

\section{Discussions} \label{sec:dis}

\subsection{Inferences from previous studies}

In this study, we conducted simulations for both HWO and LIFE, focusing on two distinct wavelength ranges corresponding to the reflected and thermal emission spectra of a modern Earth-like exoplanet. The HWO simulations covered wavelengths from 0.25 to 2~\textmu m, while the LIFE simulations spanned 4 to 18.5~\textmu m. 
Together, these missions can provide crucial insights into planetary atmospheres, their chemical compositions, and potential links to biosignature studies in exoplanetary environments. 

The gases of primary interest for biosignature studies include O$_2$, CH$_4$, CO$_2$, O$_3$, and N$_2$O, among others \citep{Schwieterman, schwieterman_Leung_2024}. On Earth, these gases are predominantly associated with biological activity, though not exclusively. Their detectability on exoplanets depends on several factors, including disequilibrium processes, atmospheric abundance, planetary distance, and observational constraints \citep{Schwieterman}. Some of these species can also be produced through abiotic processes, complicating their interpretation. Hence, identifying a molecule as a biosignature requires understanding its potential sources, encompassing both biotic and abiotic pathways.

For example, \citet{Meadows17,Meadows18} have shown that O$_2$ may not be a reliable biosignature when considered in isolation. Abiotic mechanisms such as water photolysis followed by hydrogen escape, CO$_2$ photolysis in low non-condensable gas atmospheres, or pre-main-sequence water loss around M-dwarfs can all produce substantial amounts of O$_2$ and O$_3$ without biological input. Given the limited observational data currently available for exoplanets, interpretations often rely on models guided by terrestrial analogs.

In contrast, \citet{Totton2018} emphasized that the most reliable indicators of life may arise from atmospheric chemical disequilibrium, characterized by the coexistence of strongly oxidized and reduced species that are maintained far from thermodynamic equilibrium. Their analysis indicates that, for exoplanets, the simultaneous detection of abundant CH$_4$ and CO$_2$, particularly in the absence of CO, would constitute a compelling biosignature, as maintaining high CH$_4$ levels abiotically is thermodynamically challenging. Such disequilibrium-based frameworks complement species-specific biosignature searches and can be directly tested with instruments like HWO and LIFE, which probe both CH$_4$ and CO$_2$ spectral features.

While our simulations demonstrate the detectability of certain species potentially linked to surface activity, whether biological or geological, they do not by themselves confirm the presence of biosignatures. Distinguishing between these possible origins remains a key challenge in interpreting future observations and highlights the importance of assessing atmospheric redox disequilibria alongside broader planetary context.

\subsection{Comparative detectability of key molecules with HWO and LIFE}

\begin{table*}[!ht]
	\centering
	\caption{Band-integrated signal-to-noise ratios (SNRs) for different molecular features across various wavelength bands for HWO and LIFE simulations.}
	\label{tab:Band_SNR_HWO_LIFE}
	\renewcommand{\arraystretch}{1.2}
	\begin{tabular}{|c|cccc|cc|}
		\hline
		\textbf{Models}     & \multicolumn{4}{c|}{\textbf{HWO}}                                                                                                        & \multicolumn{2}{c|}{\textbf{LIFE}}                                  \\ \hline
		\textbf{}           & \multicolumn{1}{c|}{\textbf{Mol. Feature}} & \multicolumn{3}{c|}{\textbf{Band Int. SNR (intervals)}}                         & \multicolumn{1}{c|}{\textbf{Mol. Feature}} & \textbf{Band Int. SNR} \\ \hline
		\textbf{}           & \multicolumn{1}{c|}{}                      & \multicolumn{1}{c|}{\textbf{0.15--0.5}} & \multicolumn{1}{c|}{\textbf{0.5--1}} & \textbf{1--2} & \multicolumn{1}{c|}{}                      & \textbf{4--18.5}        \\ \hline
		\textbf{Main Model} & \multicolumn{1}{c|}{$\mathrm{H_2O}$}       & \multicolumn{1}{c|}{1.39} & \multicolumn{1}{c|}{8.88} & 5.62 & \multicolumn{1}{c|}{$\mathrm{H_2O}$} & 141.88 \\ \hline
		\textbf{} & \multicolumn{1}{c|}{$\mathrm{CO_2}$} & \multicolumn{1}{c|}{1.39} & \multicolumn{1}{c|}{1.02} & 0.35 & \multicolumn{1}{c|}{$\mathrm{CO_2}$} & 52.76 \\ \hline
		\textbf{} & \multicolumn{1}{c|}{$\mathrm{O_3}$} & \multicolumn{1}{c|}{17.29} & \multicolumn{1}{c|}{3.29} & 0.3 & \multicolumn{1}{c|}{$\mathrm{O_3}$} & 11.83 \\ \hline
		\textbf{} & \multicolumn{1}{c|}{$\mathrm{O_2}$} & \multicolumn{1}{c|}{1.36} & \multicolumn{1}{c|}{4.36} & 0.38 & \multicolumn{1}{c|}{$\mathrm{N_2O}$} & 15.04 \\ \hline
		\textbf{} & \multicolumn{1}{c|}{$\mathrm{CH_4}$} & \multicolumn{1}{c|}{1.39} & \multicolumn{1}{c|}{1.02} & 0.3 & \multicolumn{1}{c|}{$\mathrm{CH_4}$} & 1.79 \\ \hline
		\textbf{Model 1} & \multicolumn{1}{c|}{$\mathrm{H_2O}$} & \multicolumn{1}{c|}{0} & \multicolumn{1}{c|}{0.06} & 0.21 & \multicolumn{1}{c|}{$\mathrm{H_2O}$} & 4.14 \\ \hline
		\textbf{} & \multicolumn{1}{c|}{$\mathrm{CO_2}$} & \multicolumn{1}{c|}{0.01} & \multicolumn{1}{c|}{0.01} & 0.27 & \multicolumn{1}{c|}{$\mathrm{CO_2}$} & 40.94 \\ \hline
		\textbf{} & \multicolumn{1}{c|}{$\mathrm{O_3}$} & \multicolumn{1}{c|}{20.33} & \multicolumn{1}{c|}{14.07} & 0 & \multicolumn{1}{c|}{$\mathrm{O_3}$} & 42.03 \\ \hline
		\textbf{} & \multicolumn{1}{c|}{$\mathrm{O_2}$} & \multicolumn{1}{c|}{1.87} & \multicolumn{1}{c|}{4.33} & 0.59 & \multicolumn{1}{c|}{$\mathrm{N_2O}$} & 0 \\ \hline
		\textbf{} & \multicolumn{1}{c|}{$\mathrm{CH_4}$} & \multicolumn{1}{c|}{0} & \multicolumn{1}{c|}{0} & 0 & \multicolumn{1}{c|}{$\mathrm{CH_4}$} & 0 \\ \hline
		\textbf{Model 2} & \multicolumn{1}{c|}{$\mathrm{H_2O}$} & \multicolumn{1}{c|}{0.04} & \multicolumn{1}{c|}{8.48} & 4.58 & \multicolumn{1}{c|}{$\mathrm{H_2O}$} & 181.96 \\ \hline
		\textbf{} & \multicolumn{1}{c|}{$\mathrm{CO_2}$} & \multicolumn{1}{c|}{0.01} & \multicolumn{1}{c|}{0} & 0.11 & \multicolumn{1}{c|}{$\mathrm{CO_2}$} & 53.28 \\ \hline
		\textbf{} & \multicolumn{1}{c|}{$\mathrm{O_3}$} & \multicolumn{1}{c|}{18.22} & \multicolumn{1}{c|}{6.46} & 0 & \multicolumn{1}{c|}{$\mathrm{O_3}$} & 23.85 \\ \hline
		\textbf{} & \multicolumn{1}{c|}{$\mathrm{O_2}$} & \multicolumn{1}{c|}{1.78} & \multicolumn{1}{c|}{4.42} & 0.58 & \multicolumn{1}{c|}{$\mathrm{N_2O}$} & 0 \\ \hline
		\textbf{} & \multicolumn{1}{c|}{$\mathrm{CH_4}$} & \multicolumn{1}{c|}{0} & \multicolumn{1}{c|}{0} & 0 & \multicolumn{1}{c|}{$\mathrm{CH_4}$} & 0 \\ \hline
		\textbf{Model 3} & \multicolumn{1}{c|}{$\mathrm{H_2O}$} & \multicolumn{1}{c|}{0} & \multicolumn{1}{c|}{0.14} & 0.33 & \multicolumn{1}{c|}{$\mathrm{H_2O}$} & 6.51 \\ \hline
		\textbf{} & \multicolumn{1}{c|}{$\mathrm{CO_2}$} & \multicolumn{1}{c|}{0.19} & \multicolumn{1}{c|}{0.09} & 0.96 & \multicolumn{1}{c|}{$\mathrm{CO_2}$} & 40.35 \\ \hline
		\textbf{} & \multicolumn{1}{c|}{$\mathrm{O_3}$} & \multicolumn{1}{c|}{17.64} & \multicolumn{1}{c|}{4.88} & 0 & \multicolumn{1}{c|}{$\mathrm{O_3}$} & 17.56 \\ \hline
		\textbf{} & \multicolumn{1}{c|}{$\mathrm{O_2}$} & \multicolumn{1}{c|}{1.72} & \multicolumn{1}{c|}{4.4} & 0.56 & \multicolumn{1}{c|}{$\mathrm{N_2O}$} & 9.12 \\ \hline
		\textbf{} & \multicolumn{1}{c|}{$\mathrm{CH_4}$} & \multicolumn{1}{c|}{0} & \multicolumn{1}{c|}{0.02} & 0.13 & \multicolumn{1}{c|}{$\mathrm{CH_4}$} & 9.13 \\ \hline
		\textbf{Model 4} & \multicolumn{1}{c|}{$\mathrm{H_2O}$} & \multicolumn{1}{c|}{0.04} & \multicolumn{1}{c|}{8.56} & 4.59 & \multicolumn{1}{c|}{$\mathrm{H_2O}$} & 141.83 \\ \hline
		\textbf{} & \multicolumn{1}{c|}{$\mathrm{CO_2}$} & \multicolumn{1}{c|}{0.01} & \multicolumn{1}{c|}{0} & 0.11 & \multicolumn{1}{c|}{$\mathrm{CO_2}$} & 54.07 \\ \hline
		\textbf{} & \multicolumn{1}{c|}{$\mathrm{O_3}$} & \multicolumn{1}{c|}{17.27} & \multicolumn{1}{c|}{2.87} & 0 & \multicolumn{1}{c|}{$\mathrm{O_3}$} & 11.78 \\ \hline
		\textbf{} & \multicolumn{1}{c|}{$\mathrm{O_2}$} & \multicolumn{1}{c|}{1.49} & \multicolumn{1}{c|}{4.23} & 0.48 & \multicolumn{1}{c|}{$\mathrm{N_2O}$} & 15.21 \\ \hline
		\textbf{} & \multicolumn{1}{c|}{$\mathrm{CH_4}$} & \multicolumn{1}{c|}{0} & \multicolumn{1}{c|}{0} & 0.02 & \multicolumn{1}{c|}{$\mathrm{CH_4}$} & 1.78 \\ \hline
		\textbf{Model 5} & \multicolumn{1}{c|}{$\mathrm{H_2O}$} & \multicolumn{1}{c|}{0.04} & \multicolumn{1}{c|}{8.63} & 4.61 & \multicolumn{1}{c|}{$\mathrm{H_2O}$} & 144.92 \\ \hline
		\textbf{} & \multicolumn{1}{c|}{$\mathrm{CO_2}$} & \multicolumn{1}{c|}{0.01} & \multicolumn{1}{c|}{0} & 0.11 & \multicolumn{1}{c|}{$\mathrm{CO_2}$} & 54.38 \\ \hline
		\textbf{} & \multicolumn{1}{c|}{$\mathrm{O_3}$} & \multicolumn{1}{c|}{17.38} & \multicolumn{1}{c|}{2.81} & 0 & \multicolumn{1}{c|}{$\mathrm{O_3}$} & 14.8 \\ \hline
		\textbf{} & \multicolumn{1}{c|}{$\mathrm{O_2}$} & \multicolumn{1}{c|}{0.65} & \multicolumn{1}{c|}{3.31} & 0.22 & \multicolumn{1}{c|}{$\mathrm{N_2O}$} & 8.47 \\ \hline
		\textbf{} & \multicolumn{1}{c|}{$\mathrm{CH_4}$} & \multicolumn{1}{c|}{0} & \multicolumn{1}{c|}{0} & 0.01 & \multicolumn{1}{c|}{$\mathrm{CH_4}$} & 2.04 \\ \hline
	\end{tabular}
\end{table*}

As a metric for detecting molecular features of interest, we define the band-integrated signal-to-noise ratio (SNR) (see Section~\ref{subsubsec:BandSNR}). Given that the exact noise characteristics of HWO and LIFE are not yet fully specified, the absolute detectability of molecular features cannot be determined with certainty. However, the band-integrated SNR values serve as a useful comparative metric, offering insights into the relative detectability of different molecules across various models. From the tabulated values of the band-integrated SNR for the Earth-like  models presented in Table \ref{tab:Band_SNR_HWO_LIFE} for HWO, we infer that ozone has very high SNR values (greater than 10 in all cases), indicating that it is observable in all scenarios. In contrast, the SNR for oxygen ranges from around 3 to 5, i.e. below our marginal detectability range and therefore not detectable under the adopted instrument and noise criteria. H$_2$O falls into the potentially detectable regime (e.g., SNR $\sim$ 8–9 in the 0.5–1 \textmu m band) in the main model as well as in models 2, 4, and 5.  However, for models 1 and 3, the SNR is too low due to the absence of water in the lower boundary (i.e., the lack of fixed surface humidity). CO$_2$ and CH$_4$, on the other hand, exhibit very low SNR values, falling below the detection sensitivity threshold of HWO under the current setup. These molecules may require significantly longer integration times to be detectable. For the Earth-like models presented in Table \ref{tab:Band_SNR_HWO_LIFE} using LIFE and the proposed integration time, we find that H$_2$O, CO$_2$, and O$_3$ have very high SNR values and would be clearly detectable in all considered models. However, CH$_4$ and N$_2$O present a more complex scenario. CH$_4$ is potentially observable in model 3, where it has a surface outgassing flux. It is not detectable when absent from the surface boundary conditions (models 1 and 2) or when H$_2$O is present at the boundary. The reason CH$_4$ is undetectable in the presence of H$_2$O at the lower boundary is that the H$_2$O absorption band in the 4–8 µm region is significantly stronger than the CH$_4$ absorption at around 7.5~\textmu m, effectively masking it. Finally, N$_2$O is detectable in the main model and in models 3, 4, and 5, where it is included in the boundary conditions. However, it remains undetectable in cases where it is absent from the boundary conditions.

For HWO, O$_3$ consistently shows the highest SNR across all models, with detectability several times greater than that of O$_2$ across the full wavelength range for the main model. H$_2$O shows potential detectability in models with a fixed surface abundance (for example, in the main model compared to Model~1, where the H$_2$O SNR decreases from 8.88 to 0.06). CH$_4$ remains undetectable in all HWO models, while CO$_2$ shows high detectability for LIFE.

For LIFE, H$_2$O is also generally the most detectable molecule, with SNR values exceeding 140 in the main model as well as in Models~4 and~5. CO$_2$ and O$_3$ remain readily observable but with lower relative SNR than H$_2$O. CH$_4$ and N$_2$O show greater variability: N$_2$O is detectable only when included through surface boundary conditions (e.g., SNR~$\sim$10 and 15 in Models~3 and~4, compared to 0 in Models~1 and~2), while CH$_4$ becomes marginally detectable (SNR~$\sim$9) in Model~3, where a surface outgassing flux is present.

These comparisons illustrate that the relative detectability of molecules depends strongly on both atmospheric composition and surface boundary conditions, highlighting which species are most likely to be observable under different planetary scenarios, independent of the precise instrument noise characteristics. This overlap in molecular detection between HWO and LIFE, with each instrument emphasizing different spectral regions and features, highlights the importance of studying these molecules collectively to obtain a more complete picture of the atmospheric composition of an Earth analog. By combining their strengths, we can better identify subtle atmospheric details and gain a deeper understanding of the planetary environment.

\subsection{Inferences on detectability from photochemical modeling under varying boundary conditions}

In the Earth-like photochemical models (see Section~\ref{sec:apndx a}), where we varied boundary conditions to assess their impact on atmospheric spectra, the VMR profiles from \texttt{VULCAN} (Figure~\ref{fig:cases_grid}), the spectra (Figures~\ref{fig:cases_spectra_HWO_grid_1} and~\ref{fig:cases_spectra_LIFE_grid_1}), and the corresponding band-integrated SNRs reveal important details. Starting with H$_2$O, its abundance increases to highly detectable levels when its surface mixing ratio is fixed. Compared to models where the surface mixing ratio of water is not fixed, this deviation is most apparent below 20~km. Therefore, the low abundance of H$_2$O and its non-detectability in both reflection and thermal emission spectra can be attributed to the absence of a fixed surface abundance and to reactions that consume H$_2$O, as described in Section~\ref{sec:apndx a}. In models lacking an H$_2$O feature in the HWO and LIFE spectra, even when bottom boundary fluxes of other molecules are included, the abundance of H$_2$O produced by atmospheric chemistry alone remains too low to be detectable. Thus, for an Earth-like detectable spectrum, the presence of H$_2$O likely indicates a fixed and relatively high surface water composition. When bottom boundary conditions are not considered, CH$_4$ also shows a major VMR shift (see Section~\ref{sec:apndx a}) and is not abundant enough to be detectable in the thermal emission spectra. This finding suggests that a constant CH$_4$ outgassing flux and the absence of surface humidity are required to raise CH$_4$ levels to observable values. In the reflection spectra, CH$_4$ has a very low SNR and would be difficult to detect with HWO under the initial conditions and proposed sensitivity considered here. In the absence of tropospheric sinks, nitrogen-bearing species such as N$_2$O can accumulate in the atmosphere from surface sources, reaching a VMR of approximately 10$^{-6}$ (see Section~\ref{sec:apndx a}), which is detectable in the thermal emission spectra. Therefore, detectable amounts of N$_2$O in the LIFE thermal emission spectra may indicate continuous surface release, making it a strong biosignature candidate \citep{Madhu25}. Thus, the spectral features of H$_2$O, CH$_4$, and N$_2$O become detectable only when there is a continuous surface supply or a stable surface abundance. Their absence in the spectra suggests that insufficient surface sources can significantly limit the detectability of these key molecules.

In all these models, O$_3$, and CO$_2$ remain abundant and detectable in the LIFE spectra, even without a specific source, sink, or fixed surface mixing ratio. The constant CO$_2$ level at 400 ppm suggests that its major destroyer, O($^1$D), and its primary producers, CO and OH, are present in such small quantities that their reactions do not significantly alter the CO$_2$ abundance. Similarly, O$_3$ is very strongly detectable by HWO for all the models however O$_2$ is barely detectable by HWO in all the models even though it accumulates in the atmosphere regardless of boundary conditions and remains stable, unaffected by the chemistry of other molecules. So O$_2$ could be detectable with HWO with higher integration time than 100 hours. However the high O$_2$ abundance, in turn, supports elevated levels of O$_3$, as ozone formation is largely driven by the availability of O$_2$.

In our combined assessment of the base model and various Earth-like scenarios with modified boundary conditions, we find that the detectability of any spectral feature is primarily determined by its abundance, which depends on both local atmospheric chemistry and surface–atmosphere interactions. For example, the detectability of O$_2$, O$_3$, and CO$_2$ is largely maintained by atmospheric chemistry, irrespective of surface boundary conditions, whereas the detectability of H$_2$O, CH$_4$, and N$_2$O strongly depends on their surface abundances. Thus, the observability of a molecular feature reflects the interplay between surface conditions and atmospheric chemistry. Furthermore, detecting a feature at a specific wavelength (e.g., CH$_2$ at $\sim$7.4~\textmu m) may indicate the absence of another molecule, pointing to different physical conditions near the surface (e.g., insufficient H$_2$O abundance). These characteristics of molecular signatures are crucial for interpreting future spectra from the HWO and LIFE missions.

In our simulations, we also identified wavelengths where CIA occurs alongside contributions from other molecules, as shown in Figures \ref{fig:HWO_final_plot} and \ref{fig:cases_spectra_HWO_grid_1} for HWO, and Figures \ref{fig:LIFE_final_plot} and \ref{fig:cases_spectra_LIFE_grid_1} for LIFE. We included all CIAs supported by PSG and discuss the specific CIAs that contribute to the spectra in Sections \ref{subsec:reflection}, \ref{subsec:thermal_emission}, and Section \ref{sec:apndx c}. While the contributions of these CIAs are minimal, removing them does not significantly impact the overall spectral characteristics. Nonetheless, the presence of these subtle CIA features provides additional context for the spectral signatures and suggests possibilities for further exploration. CIAs not discussed in this paper have no measurable impact on the spectra.

\subsection{Comparison with EPOXI observations of Earth analogs and potential limitations}

We further compared our reflected light spectra from the base modern Earth main model without clouds to the cloud-free model of \citet{Robinson_Salvador_2023} and found good agreement. However, the best-fit Earth model of \citet{Robinson_Salvador_2023}, which was used to reproduce the EPOXI observations of Earth, included a mixed-phase cloud deck (50\% liquid water and 50\% ice) with blended optical properties, specifying the cloud-top pressure, thickness, and V-band optical depth. Since our PSG-generated spectra are limited to cloud-free atmospheres, we do not compare them directly with the EPOXI observations. Moreover, PSG’s current one-dimensional cloud module does not support mixed-phase clouds within a single layer; only one cloud phase can be specified at a time, and optical properties cannot be blended within the same deck. While PSG can incorporate externally defined multi-layer or three-dimensional cloud fields, such an analysis to reproduce the EPOXI observations of Earth is beyond the scope of this paper.

\subsection{Limitations and future work}

Since the HWO mission concept and LIFE are still in the early stages of development, our simulations and assessments are based on the current understanding of their expected designs and capabilities. In this work, we are limited by the prescribed initial conditions and compositions used to simulate modern Earth-like environments. The photochemical model does not explicitly differentiate between biological and geological processes to robustly identify any species as a potential biosignature. Instead, it relies on boundary condition fluxes and deposition rates to represent them as indirect sources. Consequently, it remains challenging to designate any particular molecule as a confirmed biosignature. In the future, extending this study to develop a one-dimensional photochemical–ocean–ecosystem model would help distinguish the relative roles of biotic and abiotic processes in the origin of these molecules.

\section{Conclusions} \label{sec:con}

In this study, we perform forward modeling of several Earth analogs using temperature–pressure (T–P) profiles derived from a Numerical Weather Prediction model coupled with a one-dimensional chemical kinetics model to generate both reflection and thermal emission spectra and to evaluate their detectability under different surface conditions, which may be indirectly linked to possible biological or geological activity. Specifically, our work predicts the spectra expected for Earth analogs when observed with future facilities such as HWO and LIFE. We further assess the scientific potential of these two mission concepts for the Earth-like scenarios considered here by using the band-integrated SNR, which we classify as follows: detectable if \texttt{SNR $\geq 10$}, potentially detectable if \texttt{$5 \leq \text{SNR} < 10$}, and not detectable if \texttt{SNR $< 5$}. Our key findings and conclusions are summarized below:

\begin{itemize}

\item O$_3$ is detectable with both HWO and LIFE regardless of its surface boundary conditions for the Earth analogs, since its atmospheric abundance is primarily governed by photochemical processes. Its spectral signature remains robust even in the absence of direct surface sources or sinks.

\item For modern Earth analogs shaped by biological and geological processes, we find that the band-integrated SNR for O$_2$ in the 0.5--1.0~$\mu$m wavelength range reaches slightly above 4 for HWO. In contrast, O$_2$ features in the 0.15--0.5~$\mu$m and 1--2~$\mu$m ranges remain consistently weaker, even when surface emission or deposition processes are included for other Earth analogs. This indicates that the 0.5--1.0~$\mu$m band provides the most promising window for future O$_2$ detection, particularly with longer integration times.

\item For all Earth analog cases considered in this study, continuous surface outgassing of CH$_4$ driven by biological and geological activity, together with low surface humidity, is required for its potential detectability with LIFE. Spectral resolutions higher than 50 can be expected to further improve its detectability.

\item Similar to CH$_4$, N$_2$O becomes potentially detectable in the modeled Earth analogs by LIFE only when it is continuously supplied through surface emission fluxes associated with biological and geological activity.

\item The band-integrated SNR for H$_2$O with both HWO and LIFE increases when a fixed specific humidity is maintained at the surface, making H$_2$O potentially detectable with HWO and detectable with LIFE. This underscores the importance of surface boundary conditions for water vapor detectability.

\item CO$_2$ is consistently detectable for LIFE across nearly all Earth analogs considered here, indicating that it does not require continuous surface emission or deposition for detection and remains observable regardless of surface dynamics. However, HWO cannot detect CO$_2$ in the same modeled Earth analogs, likely due to the weaker absorption features within its accessible wavelength range.

\item Observations in the UV/VIS/NIR range with HWO provide key information about atmospheric composition and surface properties. However, for the modeled Earth analogs considered here, the band-integrated SNR indicates that neither CH$_4$ nor CO$_2$ is detectable with HWO under our assumed instrument and noise conditions.

\item Each wavelength range (UV/VIS/NIR for HWO and mid-IR for LIFE) provides unique insights along with specific observational challenges. However, combining data from these spectral regions yields a more comprehensive understanding of exoplanetary atmospheres and enhances our ability to identify Exo-Earths influenced by biological and geological processes. Such a multi-wavelength approach also improves our understanding of their role in the broader quest to search for habitable Exo-Earths.

\item The methodology and results presented here focus on the UV/VIS/NIR and mid-IR regions, in line with future mission concepts such as HWO and LIFE. However, our work and findings would also be helpful for any future missions aiming to characterize Earth-like worlds using reflected light and thermal emission spectra in these spectral regions.

\end{itemize}

\nolinenumbers


\section{acknowledgments}

L.M. sincerely thanks Dr. Ravi Kumar Kopparapu (NASA GSFC) for insightful discussions on the photochemistry of Earth analogs and on noise simulations for the Habitable Worlds Observatory (HWO). We are grateful to Dr. Vincent Kofman (NASA GSFC) for kindly sharing a template configuration file for HWO simulations in PSG and for his invaluable assistance through numerous productive discussions on PSG-based simulations. L.M. also thanks Dr. Tyler D. Robinson (University of Arizona) for providing the EPOXI observational data and the parameters used in his work, which significantly aided our discussion of the EPOXI observations of Earth. We further thank Dr. Daniel Angerhausen (ETH Zurich) for his guidance on the LIFE noise simulations. L.M. acknowledges funding support from the DAE through the NISER project RNI 4011. L.M. also thanks Dr. Vishal Gajjar (SETI Institute) for stimulating discussions on biosignatures and technosignatures, which led to the NISER–SETI collaboration. This work was supported in part by Breakthrough Listen at the University of Oxford, through a sub-award to NISER under Agreement R82799/CN013, as part of a global collaboration funded by the Breakthrough Prize Foundation. D.B.P. extends her heartfelt thanks to Swadhin Satapathy, Dwaipayan Dubey, and Rahul Arora for their valuable discussions and guidance at SEPS, NISER. O.P.J. expresses sincere gratitude to Sarin C. Jacob for his assistance. We thank the anonymous referee for constructive comments that improved this manuscript.


\vspace{5mm}
\facilities{HWO, LIFE}

\software{\texttt{PSG} \citep{Villanueva2018psg,Villanueva2022psg}, \texttt{LIFEsim} \citep{Dannert2022}, \texttt{NWP} {\url{https://nwp-saf.eumetsat.int/site/software/atmospheric-profile-data/}}, \texttt{VULCAN} \citep{Tsai_2021}, \texttt{NumPy} \citep{harris2020array}, \texttt{matplotlib} \citep{Hunter:2007}, \texttt{pandas} \citep{mckinney2010data}}

\bibliographystyle{aasjournal}
\bibliography{references}

\end{document}